\documentclass[a4paper,10pt]{article}

\usepackage[pdftex]{color,graphicx,hyperref}
\usepackage{amsmath,amsfonts,amssymb}
\usepackage{wasysym}
\usepackage{fullpage}
\usepackage{setspace}
\usepackage[labelfont=bf,labelsep=period]{caption}
\usepackage{subcaption}
\usepackage{cite}
\usepackage{booktabs}
\usepackage{multirow}
\usepackage[utf8]{inputenc}
\usepackage{color}
\usepackage{tabularx}
\usepackage[linesnumbered,ruled,vlined]{algorithm2e}

\newcommand{\eref}[1]{Eq.~(\ref{#1})}
\newcommand{\esref}[2]{Eqs.~(\ref{#1})-(\ref{#2})}
\newcommand{\sref}[1]{Section~\ref{#1}}
\newcommand{\fref}[1]{Fig.~\ref{#1}}
\newcommand{\tref}[1]{Table~\ref{#1}}

\usepackage{authblk}

\title{\textbf{Reentrant phase transitions in threshold driven contagion on multiplex networks}}
\author[1]{Samuel Unicomb}
\author[2]{Gerardo I\~{n}iguez} 
\author[2]{J\'{a}nos Kert\'{e}sz}
\author[1,*]{M\'{a}rton Karsai}
\affil[1]{Universit\'{e} de Lyon, ENS de Lyon, INRIA, CNRS, UMR 5668, IXXI, 69364 Lyon, France}
\affil[2]{Department of Network and Data Science, Central European University, H-1051 Budapest, Hungary}
\affil[3]{Department of Computer Science, Aalto University School of Science, 00076 Aalto, Finland}
\affil[4]{IIMAS, Universidad Nacional Auton\'{o}ma de M\'{e}xico, 01000 Ciudad de M\'{e}xico, Mexico}
\affil[*]{Corresponding author: marton.karsai@ens-lyon.fr}
\date{}                     
\begin{document}

\maketitle

\begin{abstract}
Models of threshold driven contagion explain the cascading spread of information, behavior, systemic risk, and epidemics on social, financial and biological networks. At odds with empirical observation, these models predict that single-layer unweighted networks become resistant to global cascades after reaching sufficient connectivity. We investigate threshold driven contagion on weight heterogeneous multiplex networks and show that they can remain susceptible to global cascades at any level of connectivity, and with increasing edge density pass through alternating phases of stability and instability in the form of reentrant phase transitions of contagion. Our results provide a novel theoretical explanation for the observation of large scale contagion in highly connected but heterogeneous networks.
\end{abstract}

\section*{Introduction}

Information-communication technology has radically transformed social and economic interaction~\cite{borge2013cascading}, introducing new means of transmitting ideas, behavior, and innovation~\cite{rogers2010diffusion, centola2007complex}, overcoming limitations imposed by time and cognitive constraints~\cite{dunbar1992neocortex, goncalves2011modeling}. The same technology provides an increasingly accurate picture of human interaction, mapping the underlying network structures that mediate dynamical processes, like epidemics~\cite{joh2009dynamics,takaguchi2013bursty}. In complex contagion~\cite{centola2007complex}, characteristic of the spreading of innovation, rumors, or systemic risk, transmission is a collective phenomenon in which all social ties of an individual may be involved. Node degree, or number of links, is therefore critical to the dynamical outcome~\cite{watts2002simple}; large relative neighbor influence is easier to achieve the smaller the ego network. This behavior is well captured by threshold models of social contagion on single-layer unweighted networks, which predict large-scale cascades of adoption in relatively sparse networks~\cite{granovetter1978threshold,watts2002simple,ruan2015kinetics,karampourniotis2015impact,ruan2015kinetics}. In empirical social networks, however, individuals can maintain hundreds of ties~\cite{goncalves2011modeling,dunbar2012social}, with interaction strength varying across social contexts~\cite{jo2018stylized,unicomb2018threshold,burkholz2016damage}, yet still exhibit frequent system-wide cascades of social contagion~\cite{karsai2016local,bakshy2011everyone,ugander2012structural,dow2013anatomy,gleeson2017temporal}.

We address this issue by incorporating relevant features of empirical social networks into a conventional threshold model. We consider that network ties are heterogeneous, and can be characterized by edge ``types''. In the case of social networks, these edge types vary in ``quality''~\cite{granovetter1977strength,kawachi2001social}, usually associated with the intimacy or perceived importance of a relationship between individuals~\cite{zhou2005discrete}, and scale with the strength of interpersonal influence~\cite{cialdini2004social,turner1991social}. Heterogeneity in tie quality is well modeled by multiplex structures, as has been recognized in both network ~\cite{kivela2014multilayer,boccaletti2014structure} and social science~\cite{bruce1969norms,verbrugge1979multiplexity}, particularly regarding social contagion~\cite{yagan2012analysis,brummitt2012multiplexity,lee2014threshold,zhuang2017clustering}. In multiplex models of social networks, individual layers represent the social context of a relationship (e.g. kinship, acquaintance), allowing us to classify ties by social closeness, as recognised by Dunbar's intimacy circle theory~\cite{zhou2005discrete}. According to this theory, due to cognitive and time resources being finite but necessary to maintaining social ties, individuals actively cultivate a limited number of relationships, organising them into intimacy circles that increase in size as they decrease in importance. Ego networks thus comprise a small but high-intimacy circle of close relationships, like family and long term friends, followed by large but low-intimacy circles of distant friends and acquaintances. Empirical evidence shows the distribution of dyadic social commitments (number of interactions or time devoted to peers) to be strongly heterogeneous~\cite{marsden1984measuring,onnela2007structure}. Strikingly, this inverse relation between the cost of maintaining an edge type, and the abundance of that edge type, can be seen as an entropy maximisation process~\cite{tamarit2018cognitive} that applies to any system with heterogeneous cost of edge formation and finite node resources. As such, although we use the language of social networks, our results are of relevance to other systems, e.g., financial~\cite{elliott2014financial, battiston2009liaisons,amini2016resilience} and biological~\cite{joh2009dynamics,takaguchi2013bursty} contagion. 

Using analytical and numerical tools, we show that layer hierarchy can lead to global cascades in multiplexes with average degree in the hundreds or thousands, perturbed by a single initial adoption. We report the novel observation that in a multiplex network with increasing link density a sequence of phase transitions occur, resulting in alternating phases of stability and instability to global cascades.

\section*{Results}

\begin{figure}[t]
\centering
\hspace*{-1.5mm}
\includegraphics[scale=1]{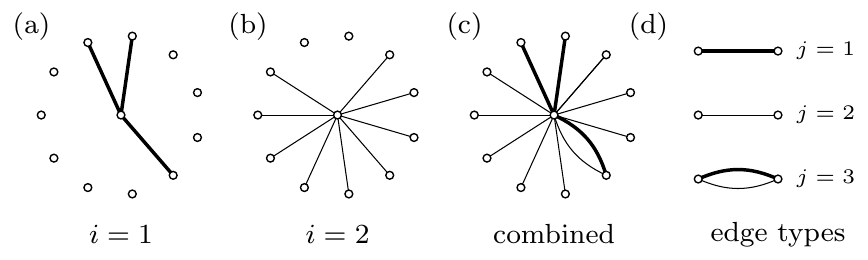}
\caption{(a-b) Egocentric view of multiplex structure with $M = 2$ layers, where edge density increases ($\delta_{z} > 1$) and edge weight decreases ($\delta_{w} < 1$) in each layer $i$. (c) Egocentric network overlap between layers. (d) Emergent edge types in the overlapping network. In the multiplex, the central node has degree vector $\textbf{k} = (2,8,1)^{T}$, encoding layer overlap.
\label{fig:fig1}}
\end{figure}

Our model builds upon previous studies of threshold driven processes~\cite{granovetter1978threshold,watts2002simple,ruan2015kinetics,karampourniotis2015impact} and multiplex networks~\cite{kivela2014multilayer,boccaletti2014structure}. We define contagion as a binary-state dynamics over a weighted, undirected multiplex network of $N$ nodes connected throughout $M$ layers (\fref{fig:fig1}). A node represents an individual $u$, and layer $i$ the social context in which individuals interact, $1 \leq i \leq M$. The degree of $u$ in each layer $i$ takes discrete values $k_{i} = 0, \ldots, N - 1$ according to the degree distribution $P_{i}(k)$. Edge weights $w_{i}(u, v)$ follow the continuous distribution $P_{i}(w)$ and capture the total capacity of nodes $u$ and $v$ to influence each other via layer $i$. The network allows for layer overlap~\cite{cellai2013percolation} as nodes may be connected in multiple layers, modeling individuals who share several social contexts [\fref{fig:fig1}(c)]. 
For simplicity, we assume that node degree is independent across layers, and that degree and weight distributions $P_{i}(k)$ and $P_{i}(w)$ differ by layer only in their means $z_i = \sum_k k P_{i}(k)$ and $w_i = \int w P_{i}(w)dw$, otherwise retaining their functional form. In order to reproduce the hierarchical organization of edges suggested by intimacy circle theory~\cite{zhou2005discrete}, we assume that the mean degree $z_i$ and weight $w_i$ scale with layer index $i$ as
\begin{equation}
z_{i + 1} = \delta_{z}z_{i} \hspace{4mm} \text{and} \hspace{4mm} w_{i + 1} = \delta_{w}w_{i},
\label{eqn:delta}
\end{equation}
with $\delta_{z} \geq 1$ and $\delta_{w} \leq 1$. In other words, ego networks comprise a small number of high-intimacy neighbors [\fref{fig:fig1}(a)] and a larger number of low-intimacy neighbors [\fref{fig:fig1}(b)]. We fix the average total degree $z = \sum_i z_i$ as well as $\delta_{z}$, which determines $z_{i}$. We also impose the arbitrary constraint $\langle w \rangle = 1$ and fix $\delta_{w}$, which determines $w_{i}$ (see Supplementary Information [SI]).

In a binary-state model of contagion, nodes are in one of two mutually exclusive states, susceptible or infected (also called adopter or activated in the social contagion literature). Since nodes must be either connected or disconnected via each of the $M$ network layers, their interaction is characterized by one of $2^{M}-1$ resultant edge types [\fref{fig:fig1}(d)], disregarding nodes disconnected in all layers, and indexing by $j$ such that $1 \leq j \leq 2^{M} - 1$. Node configuration is thus described by the number of neighbors $k_{j}$ and infected neighbors $m_{j}$ across edges of type $j$, with $0 \leq m_{j} \leq k_{j}$. We store $k_{j}$ and $m_{j}$ in the degree vector $\textbf{k}$ and partial degree vector $\textbf{m}$, respectively (of dimension $2^{M} - 1$). Note that we consistently index layer by $i$ and resultant edge type by $j$.

\begin{table}
\centering
\noindent\begin{tabular}{@{}*{4}{p{.25\textwidth}@{}}}
    \hline\hline
	    weighted sum  & multiplex \textit{or} & multiplex \textit{and}\\
	  	\hline 
        $q_{m} \geq \phi q_{k}$
	    \hspace{0mm}
	    &
	   	$\exists i \hspace{2mm}\text{s.t.}$
	   	\hspace{0mm} 
	   	$q_{m_{i}} \geq \phi_{i} q_{k_{i}}$
	  	\hspace*{0mm}
	    &
	    $q_{m_{i}} \geq \phi_{i} q_{k_{i}}
	    \hspace{2mm} 
	    \forall i$\\
    \hline\hline
\end{tabular}
  \caption{Extensions of the Watts threshold rule to multiplex networks. Node state is determined by a single threshold $\phi$ and a weighted sum of influence over layers, or by individual layer thresholds $\phi_{i}$ and influence within each layer. In the former the multiplex can be projected to a single weighted layer without loss of information relevant to the dynamics. \label{tab:tab1}}
\end{table}

The threshold rule proposed by Watts~\cite{granovetter1978threshold, watts2002simple, ruan2015kinetics, karampourniotis2015impact} defines the fraction $\phi$ of neighbors that must be infected for a susceptible ego to adopt. This rule can be extended to multiplex networks in several ways (\tref{tab:tab1}). Denoting the set of neighbors of node $u$ in layer $i$ by $\mathcal{N}_{i}(u)$, the total influence upon $u$ in layer $i$ is $q_{k_{i}} = \sum_{v\in \mathcal{N}_{i}(u)}w_{i}(u,v)$. Restricted to infected neighbors, $\mathcal{N}_{i}(u)|_{I}$, this gives $q_{m_{i}} = \sum_{v\in \mathcal{N}_{i}(u)|_{I}}w_{i}(u,v)$. 
In one variant of the threshold rule, nodes perceive influence in aggregate, summed over layers (reminiscent of neural networks \cite{gerstner2014neuronal,iyer2013influence}) and adopt with respect to a single threshold if $q_{m} \geq \phi q_{k}$, where $q_{k} = \sum_{i}q_{k_{i}}$ and $q_{m} = \sum_{i}q_{m_{i}}$ (weighted sum rule).
In another variant, node state is determined by $M$ layer thresholds $\phi_{i}$, along with influence $q_{k_{i}}$ and $q_{m_{i}}$ within layers. A node activates when $q_{m_{i}} \geq \phi_{i}q_{k_{i}}$ in every layer (multiplex \textit{and} rule by Lee~\cite{lee2014threshold}), or in at least one layer (multiplex \textit{or} rule~\cite{lee2014threshold}). Our aim is to show that multiplex networks following the structure of intimacy circle theory exhibit reentrant phase transitions for both the weighted sum and the multiplex \textit{or} threshold rules. Note that if weights are uniform within each layer and node state is determined by decisions within layers (\textit{and} and \textit{or} rules), then the structure is effectively unweighted. We show that even with this loss of weight information, reentrant phase transitions can still emerge due to contagion within layers.

\begin{figure}[h!]
\centering
\hspace*{-2mm}
\vspace*{-4mm}
\includegraphics[scale=1]{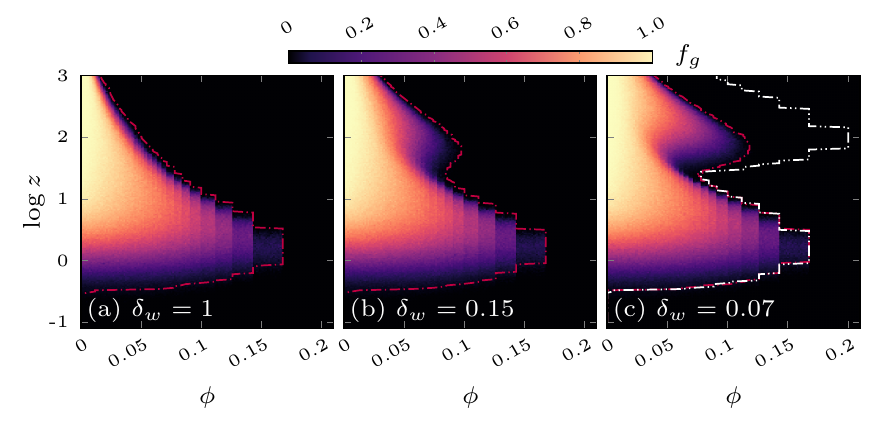}
\caption{Emergence of a high-$z$ cascading phase in $(\phi, z)$-space for the weighted sum rule, for LN degree distribution, fixed $\delta_{z} = 50$, $\gamma = 0.5$ and decreasing $\delta_{w}$. MC simulations provide the relative frequency $f_{g}$ of global cascades, after $10^{3}$ instances of single node perturbation, in a configuration-model multiplex with $N = 10^6$. In (a) we recover the classic Watts phase diagram ($\delta_{w} = 1$). The constraint $\langle w \rangle = 1$ means $\textbf{w} = (1, 1)^{T}$, $(6, 0.9)^{T}$ and $(11, 0.8)^{T}$, from (a) to (c). The outer contour (dash-double dotted white line) in (c) shows the case $\delta_{w} \rightarrow 0$ [$\delta_w =10^{-3} $; see heat map in \fref{fig:fig3}(a)]. Dash-dotted red lines show agreement with LSA prediction.}
\label{fig:fig2}
\end{figure}

\begin{figure}
\centering
\hspace*{-2mm}
\vspace*{-4mm}
\includegraphics[scale=1]{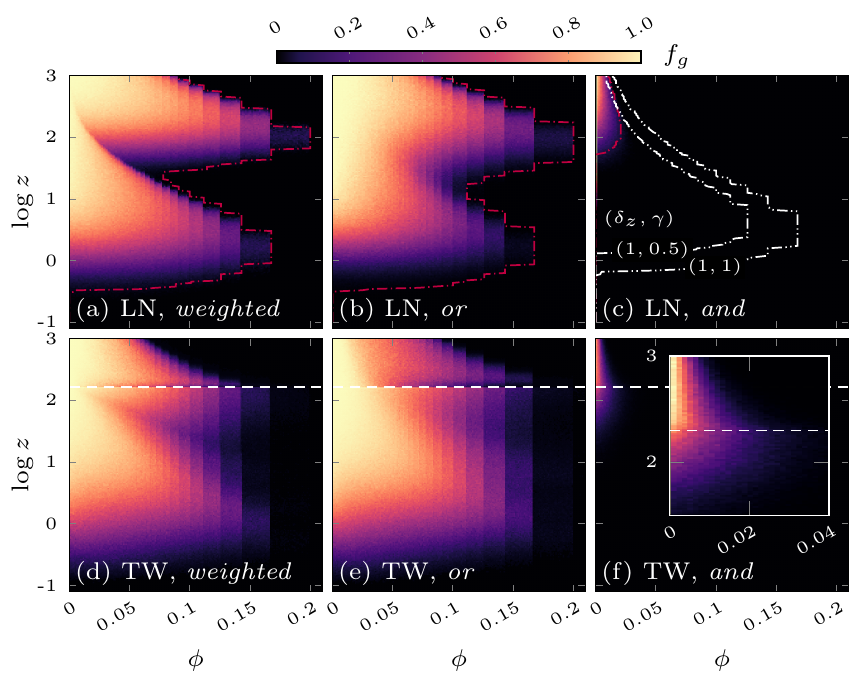}
\caption{Relative frequency $f_{g}$ of global cascades in LN (top) and TW (bottom) multiplexes with $M = 2$ layers. LN networks in (a-c) are synthetic (standard deviation $\sigma_{k_{i}} = 2z_{i}$, overlap $\gamma = 0.5$, and density scaling $\delta_{z} = 50$).  (a)  Maximal weight heterogeneity ($\delta_{w} = 10^{-3}$) leads to reentrant transitions in the weighted sum rule. (b) Reentrant phase transitions also appear for the \textit{or} threshold rule. (c) Under the \textit{and} rule only one global cascading phase emerges, which vanishes when $\gamma = 0$. Decreasing $\delta_{z}$ and increasing $\gamma$ expands the region of susceptibility to global cascades. See the outer dash-double dotted white contours (the LSA solution for $\delta_{z} = 1$, with $\gamma = 0.5$ and $1$). (d-f) Reentrant phase transitions under the weighted sum and \textit{or} rules in an empirical Twitter network ($\delta_{z} = 30.2$ and $\gamma = 0.45$). The dashed horizontal line at $z = 166$ is the empirical density, with sparsification providing lower $z$ values, and densification higher $z$ (see SI). (f) A single phase region observed in the \textit{and} multiplex rule. LN and TW networks have size $N = 10^{5}$ and $N = 3.7 \times 10^{5}$. We obtain $f_{g}$ via $10^{3}$ realisations of single node perturbation. Dash-dotted red lines show the LSA prediction.
}
\label{fig:fig3}
\end{figure}

\begin{figure}[t]
\centering
\hspace*{-2.0mm}
\vspace*{-2mm}
\includegraphics[scale=1]{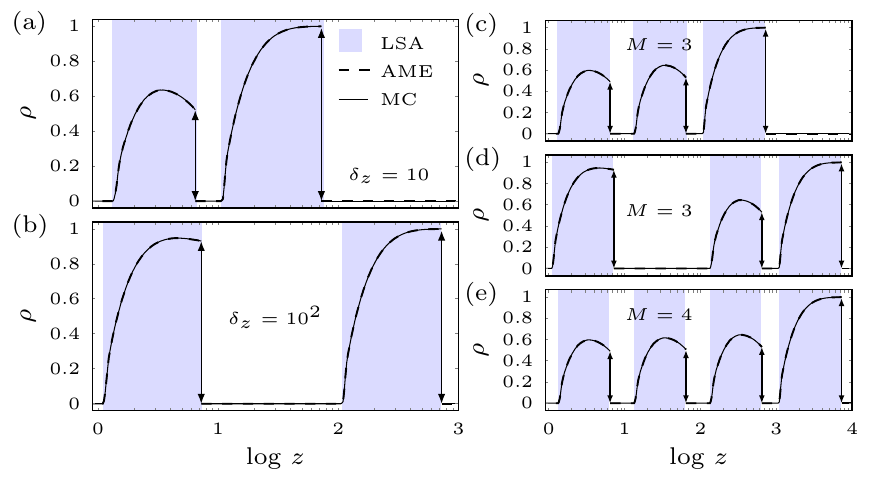}
\caption{Steady state global cascade size as a function of average degree $z$, for constant threshold $\phi = 0.15$ and maximal weight heterogeneity ($\delta_{w} = 10^{-6}$), using the weighted sum rule. Degree distributions are Poisson and overlap is $\gamma = 0$. Shaded intervals due to LSA indicate systems with a positive leading eigenvalue (see SI); dashed lines indicate the steady state solution of \eref{eqn:mastereqn}; and MC solutions are given by the solid curve (error bars narrower than line width). (a, b) Increasing density skewness $\delta_{z}$ delays the onset of high-$z$ phases of contagion, and allows larger cascades in low-$z$ phases in an $M = 2$ layer multiplex. (c-e) Increasing the number of layers to $M=3$ and $4$ induces $6$ and $8$ phase transitions in cascade size, respectively. (d) Varying $\delta_{z}$, such that $z_{2} / z_{1} = 100$, and $z_{3} / z_{2} = 10$. MC results are averaged over $10^{3}$ realizations of single node perturbation, with $N = 10^{7}$.}
\label{fig:fig4}
\end{figure}

We solve for our model using the approximate master equation (AME) formalism~\cite{gleeson2011high,gleeson2013binary}. Similar to earlier solutions~\cite{ruan2015kinetics,karsai2016local,unicomb2018threshold}, at time $t$, the density of infected nodes $\rho$ and the average probability $\nu_{j}$ that a $j$-type neighbor of a susceptible node is infected are governed by the system of coupled differential equations,
\begin{eqnarray}
\begin{aligned}
\dot{\nu}_{j} &= g_{j}(\boldsymbol{\nu}, t) - \nu_{j}, \\
\dot{\rho} &= h(\boldsymbol{\nu}, t) - \rho,
\end{aligned}
\label{eqn:mastereqn}
\end{eqnarray}
where $g_{j}(\boldsymbol{\nu}, t)$ and $h(\boldsymbol{\nu}, t)$ are known functions (see SI and ~\cite{unicomb2018threshold}). A numerical solution of \eref{eqn:mastereqn} provides the dynamical evolution of each threshold rule, and linear stability analysis (LSA)~\cite{porter2016dynamical} the region in $(\phi, z)$-space allowing global cascades (dash-dotted lines in Figs.~\ref{fig:fig2} and \ref{fig:fig3}; shaded intervals in Fig.~\ref{fig:fig4}) (further details in SI). We derive a global cascade condition via the Jacobian matrix $\mathbf{J}$ corresponding to \eref{eqn:mastereqn}, evaluated at the fixed point $\boldsymbol{\nu}^{*} = \boldsymbol{0}$,
\begin{eqnarray}
\begin{aligned}
J_{ij}^{*} &= -\delta_{ij} + \dfrac{\partial g_{i} (\boldsymbol{\nu})}{\partial \nu_{j}}\Bigr|_{\substack{\boldsymbol{\nu} = \boldsymbol{\nu}^{*}}},\\
\end{aligned}
\label{eqn:stability}
\end{eqnarray}
which has eigenvalues $\lambda_{j}$. Global cascades occur if $\text{Re} (\lambda_{j}) > 0$ for any $j = 1, \ldots, 2^M-1$. In what follows we study the response of the network to an infinitesimal perturbation, or single infected seed, and record the relative frequency $f_{g}$ of global cascades via Monte Carlo (MC) simulations. Regions in $(\phi, z)$-space with non-zero $f_{g}$ in the $N \to \infty$ limit are well predicted by the spectrum of \eref{eqn:stability}. For simplicity we assume uniform edge weights with value $w_{i}$ within layers, which can be easily generalised (see SI).

The weighted sum rule leads to a high-$z$ cascading phase, and thus reentrant phase transitions for constant $\phi$, in an $M = 2$ layer multiplex with a log-normal (LN) degree distribution in each layer (\fref{fig:fig2}, distribution details in SI). In two layers, we define layer overlap as $\gamma = |E_{1} \cap E_{2}| / |E_{1}|$, where $E_{i}$ is the edge set in layer $i=1, 2$ ($|E_{1}| < |E_{2}|$). We can increase weight heterogeneity by decreasing the weight scaling factor $\delta_{w}$, resulting in a second cascading regime. As explained in~\cite{watts2002simple}, global cascades are due to ``vulnerable'' nodes with sufficiently low threshold so that a single neighbor can infect them. A cascading phase is formed in $(\phi, z)$-space when vulnerable nodes form a percolating cluster. In single-layer unweighted networks, large $z$ results in most nodes being stable against neighbor infection, and cascades becoming exponentially rare. However, under the weighted sum rule, weight heterogeneity allows one high-influence infected neighbor to dominate a node's total received influence if remaining neighbors have low influence. Crucially, such configurations are abundant when the conditions $\delta_{z} > 1$ and $\delta_{w} < 1$ are satisfied simultaneously, resulting in a percolating vulnerable cluster at high $z$. In the low-$z$ phase, cascades are mediated by the connectivity of the weak layer, since the strong layer is too sparse to percolate. In the high-$z$ phase, strong edges percolate and determine the stability of adjacent nodes that are otherwise stable to the dense weak layer. Both regions are accurately predicted by LSA [see \fref{fig:fig2} and velocity field analysis of \eref{eqn:mastereqn} in SI]. Note that other mechanisms are able to generate additional transitions in $(\phi, z)$-space (e.g., degree assortativity in ~\cite{zhuang2017clustering}).

We compare the behavior induced by the threshold rules of \tref{tab:tab1} for configuration-model multiplexes with LN degree distributions and a real-world multiplex extracted from Twitter (TW) (\fref{fig:fig3}). TW comprises a sparse, strongly interacting layer ($z_{1} = 5.4$) formed by mutual-mention interactions between $N = 3.7 \times 10^{5}$ users, and a dense layer of weak links ($z_{2} = 163$) formed by the follower network of the same users. The two layers (taken as undirected; data details in SI) exhibit an overlap $\gamma = 0.45$. In order to explore the effect of single node perturbation over $(\phi, z)$-space, we remove edges uniformly at random from TW, decreasing its average degree $z$ below its observed value of $165.8$ [dashed lines in \fref{fig:fig3}(d-f)]. Conversely, we use a model of network densification known as the Forest-Fire process~\cite{leskovec2007graph} to extrapolate to higher $z$ values (details in SI).

Assuming the weighted sum threshold rule [\fref{fig:fig3}(a) and (d)], we find reentrant cascading phases under maximal weight 
heterogeneity ($\delta_{w} = 10^{-3}$) [for the approach to maximal heterogeneity see \fref{fig:fig2}(a-c) for LN, and SI for TW]. The multiplex \textit{or} condition also leads to reentrant transitions in both LN and TW networks [\fref{fig:fig3}(b) and (e)]. The onset of the high-$z$ cascading phase, and thus of the reentrant transition, is triggered by the structural percolation of the sparse layer. Since the \textit{or} rule considers influence within layers, and $P_{i}(w)$ is uniform here, the structure is effectively unweighted, underlining that density skewness is sufficient to trigger a reentrant phase when thresholds are layered. For both LN and TW networks, overlap $\gamma$ and density skewness $\delta_{z}$ determine the stability under the \textit{and} threshold rule [\fref{fig:fig3}(c) and (f)]. Being the most restrictive condition, the \textit{and} rule suppresses reentrant phase transitions and confines global cascades to a single phase at low $\phi$, with cascades vanishing when $\gamma = 0$. As $\delta_{z}$ decreases and $\gamma$ increases [\fref{fig:fig3}(c)], overlapping edges, necessary for mediating cascades under the \textit{and} rule, become more abundant and increase the area of the unstable phase [\fref{fig:fig3}(c)]. For simplicity, we set $\phi_{i} = \phi$ for the \textit{and} and \textit{or} rules. 
Inspection of the contours of \fref{fig:fig3}(a-c) reveals that the weighted sum rule occupies an area intermediate between the \textit{and} and \textit{or} rules; we perform a comparative eigenvalue analysis in the SI to argue that this is generally the case.

We illustrate using the weighted sum rule that density skewness $\delta_{z}$ determines the average degree $z$ at which reentrant phases are triggered [\fref{fig:fig4}(a) and (b)]. This is because the structural percolation transition of individual layers is necessary for the percolation of a subgraph of vulnerable nodes; the value of $z$ at which this occurs depends on $\delta_{z}$. Increasing the number of layers in the network also creates additional phases of contagion [see \fref{fig:fig4}(c-e) for $M=3, 4$]. When $\delta_{z}$ differs between layers, the onset of contagion phases may be delayed or promoted [\fref{fig:fig4}(d)]. In lower phases, strong edges that are too sparse to percolate structurally inhibit cascades driven by edges that are denser but weaker, leading to ``partial" cascades that are global but do not fill the network [e.g., lower phase in \fref{fig:fig4}(a)]. This is due to the immunizing effect of strong edges in information diffusion; pairs of susceptible nodes connected by a sufficiently strong edge are impossible to infect if all other neighbors are weak, even if all those weak neighbors are infected. These configurations are abundant when the strong layer is yet to undergo structural percolation.

Our results demonstrate that global information cascades emerge in arbitrarily dense networked systems, typically viewed as stable against small perturbations. The types of multiplex structure triggering this behavior are elementary, and have even been derived from an entropy maximisation process. We have shown that skewness in edge density by layer is necessary for the emergence of reentrant phase transitions under all variants of the threshold rule, but sufficient only when thresholds are layered and the \textit{or} rule applied. When influence is summed over layers and evaluated with respect to a single threshold, an additional weight skewness condition is necessary. We confirm these phenomena using an analytical formalism that we have extended to multiplex networks, as well as simulation, both on synthetic networks and an empirical Twitter multiplex where all results are recovered. Our results suggest approaches to network design that may promote or suppress system-wide cascades of threshold driven contagion.

\textbf{Acknowledgements:} 
We acknowledge the P\^{o}le Scientifique de Mod\'{e}lisation Num\'{e}rique (and L. Taulelle for technical assistance) from ENS Lyon for their computing support; D. Knipl for support in the initial stages of the project; anonymous referees for deep engagement with our work; the ACADEMICS grant of IDEXLYON, Univ. Lyon, PIA (ANR-16-IDEX-0005); SoSweet (ANR-15-CE38- 0011), and MOTIf (18-STIC-07) projects.

\newpage
\begin{center}
{\Large \textbf{SUPPLEMENTARY INFORMATION}}
\end{center}

\section{ Analytic solution}

\subsection{Reduced-dimension approximate master equations}

In this work we extend the edge-heterogeneous, approximate master equation  (AME) formalism (first presented in \cite{unicomb2018threshold} and described in detail in its Supplementary Information) to multiplex networks  comprised of $M$ layers. This formalism is configuration based, meaning that we solve for the densities of susceptible  and infected nodes over time $t$ according to their local configurations  of degree and infected neighbours, denoted $(\textbf{k}, \textbf{m})$. In the lowest level of its formulation, we solve the  AMEs for all densities of each class, $s_{\textbf{k}, \textbf{m}}$ and $i_{\textbf{k}, \textbf{m}}$, or the fraction of susceptible and infected nodes, respectively, with degree vector $\textbf{k} = (k_1, \ldots, k_{2^M-1})$ that have partial degree vector $\textbf{m} = (m_1, \ldots, m_{2^M-1})$ at time $t$,  where the index $j = 1, \ldots, 2^M-1$ runs over composite edge type. To ensure a finite space of such configurations, we require discrete edge types, and therefore a discrete set of weights. For simplicity, we assume a uniform weight distribution with each layer, such that each edge type is associated with a single weight. It is straightforward to relax this assumption to general discrete weight distributions, but at the cost of computational complexity. The master equation determining $s_{\textbf{k}, \textbf{m}}$ at time $t$, for a monotone dynamics such as complex contagion where recovery from the infected state is impossible, is given by
\begin{equation}
    \dfrac{d}{dt}s_{\textbf{k}, \textbf{m}} = -F_{\textbf{k}, \textbf{m}}s_{\textbf{k}, \textbf{m}} - \sum_{j = 1}^{2^{M} - 1}\beta_{j}^{s}(k_{j} - m_{j})s_{\textbf{k}, \textbf{m}} + \sum_{j = 1}^{2^{M} - 1}\beta_{j}^{s}(k_{j} - m_{j} + 1)s_{\textbf{k}, \textbf{m} - \textbf{e}_{j}}, 
\end{equation}
where $\textbf{e}_{j}$ is the $j$-th basis vector, and $\beta_{j}^{s}$ the rate of $j$-type neighbour infection. More precisely, it is the rate at which $j$-type susceptible neighbours of susceptible nodes become infected, averaged across the entire configuration space. It is defined as
\begin{equation}
    \beta_{j}^{s} = \dfrac{\sum_{\textbf{k}, \textbf{m}}P(\textbf{k})(k_{j} - m_{j})F_{\textbf{k}, \textbf{m}}s_{\textbf{k}, \textbf{m}}}{\sum_{\textbf{k}, \textbf{m}}P(\textbf{k})(k_{j} - m_{j})s_{\textbf{k}, \textbf{m}}},
\end{equation}
with sums being over all $(\textbf{k}, \textbf{m})$ defined in the system. Finally, $F_{\textbf{k}, \textbf{m}}dt$ is the probability that a node with configuration $(\textbf{k}, \textbf{m})$ adopts over an interval $dt$. In the case of complex contagion, the AMEs can be reduced in dimension to the system
\begin{subequations}
\label{eq:reducedAMEs}
\begin{align}
\dot{\nu}_{j} &= g_j(\boldsymbol{\nu}) - \nu_{j}, \\
\dot{\rho} &= h(\boldsymbol{\nu}) - \rho,
\end{align}
\end{subequations}
where the functions $g_{j}(\boldsymbol{\nu})$ and $h(\boldsymbol{\nu})$ are defined by
\begin{equation}
\label{eqn:reduced_g}
g_{j}(\boldsymbol{\nu}) = \sum_{\textbf{k}}\dfrac{k_{j}}{c_{j}}P(\textbf{k})\sum_{\textbf{m}}f(\textbf{k},\textbf{m})B_{k_{j} - 1,m_{j}}(\nu_{j})\prod_{i \neq j}^{2^M-1} B_{k_{i}, m_{i}}(\nu_{i})
\end{equation}
and
\begin{equation}
\label{eqn:reduced_h}
h(\boldsymbol{\nu}) =\sum_{\textbf{k}}P(\textbf{k})\sum_{\textbf{m}}f(\textbf{k},\textbf{m})\prod_{j = 1}^{2^M-1}B_{k_{j}, m_{j}}(\nu_{j}),
\end{equation}
with $B_{k_{i}, m_{i}}(\nu_{i})$ the binomial distribution.  The function $f(\textbf{k},\textbf{m})$, implementing the response of a node with degree vector $\textbf{k}$ to a set of infected neighbours encoded by $\textbf{m}$, is equal to 1 if one of the conditions in Table I of the main text is satisfied, and 0 otherwise (see \sref{ssec:M2case} for explicit expressions of the response function for $M=2$ in all multiplex threshold rules explored here).
	
In \esref{eqn:reduced_g}{eqn:reduced_h}, $P(\textbf{k})$ is the probability that a randomly selected node has degree vector $\textbf{k}$. Given that our multiplex  network is maximally random up to the degree distribution  $P_j(k_j)$ of each edge type $j = 1, \ldots, 2^M-1$, and that the  corresponding degrees $k_j$ are uncorrelated, $P(\textbf{k})$  is the product of all edge-type degree probabilities,
\begin{equation}
\label{eq:VecDegDist}
P(\textbf{k}) = \prod_{j=1}^{2^M-1} P_j(k_j),
\end{equation}
with $c_j = \sum_{k_j} k_j P_j(k_j)$ the average degree for edge type $j$. If $G_j(u) = \sum_{k_j} P_j(k_j) u^{k_j}$ is the probability-generating function associated with edge type $j$, then the aggregate degree $k = \sum_j k_j$ has probability-generating function $G(u) = \sum_k P(k) u^{k} = \prod_j G_j(u)$, from which the aggregate degree distribution $P(k)$ can be obtained.

\subsection{Cascade condition}

We can also use the AME formalism to derive a cascade condition, as has been done previously for the Watts model~\cite{porter2016dynamical} and for  complex contagion in unweighted networks~\cite{ruan2015kinetics}.
We perform a linear stability analysis of the reduced AME system in \eref{eq:reducedAMEs} around the fixed point $(\boldsymbol{\nu}^*, \rho^*) = (\boldsymbol{0}, 0)$, corresponding to a total lack of infection. If $(\boldsymbol{\nu}^*, \rho^*)$ is unstable, then any small perturbation (like a single infected node at $t = 0$) can drive the system out of equilibrium and create a global cascade of infection where $\rho > 0$, that is, a system where a non-vanishing fraction of nodes is infected in the limit $N \to \infty$. Since the system $\dot{\nu}_{j} = g_j(\boldsymbol{\nu}) - \nu_{j}$ is closed, the stability of \eref{eq:reducedAMEs} is determined by the stability of this equation at $\boldsymbol{\nu}^* = \boldsymbol{0}$. According to linear stability theory, a local instability exists if the Jacobian matrix of the system evaluated at the fixed point,
\begin{equation}
\label{eq:jacobMat}
J^*_{ji} = -\delta_{ji} + \frac{\partial g_j(\boldsymbol{\nu})}{\partial \nu_{i}} \bigg|_{\boldsymbol{\nu} = \boldsymbol{\nu}^*},
\end{equation}
has at least one eigenvalue with a real part larger than zero. We can write the partial derivative in \eref{eq:jacobMat} explicitly by considering the expansion of  $B_{k_{i}, m_{i}}(\nu_{i})$ in Eq.~\ref{eqn:reduced_g}, 
\begin{equation}
\label{eq:derivGfactor}
\frac{\partial g_j}{\partial \nu_{i}} =
\begin{cases}
\sum_{\textbf{k},\textbf{m}} \frac{k_j}{c_j}  P(\textbf{k}) f(\textbf{k}, \textbf{m}) \dot{B}_{k_{j} - 1,m_{j}}(\nu_{j}) \prod_{i \neq j} B_{k_{i}, m_{i}}(\nu_{i}) \quad & j = i\\

\sum_{\textbf{k},\textbf{m}} \frac{k_j}{c_j}  P(\textbf{k}) f(\textbf{k}, \textbf{m}) B_{k_{j} - 1,m_{j}}(\nu_{j}) \dot{B}_{k_{i} , m_{i}} (\nu_{i}) \prod_{l \neq j,i} B_{k_{l}, m_{l}} (\nu_{l}) \quad & j \neq i
\end{cases},
\end{equation}

where
\begin{equation}
\label{eq:derivBinom}
\dot{B}_{k_{j} - 1,m_{j}}(\nu_{j}) = \binom{k_j - 1}{m_j} \big[ m_j \nu_{j}^{m_j - 1} (1 - \nu_{j})^{k_j - 1 - m_j} - (k_j - 1 - m_j) \nu_{j}^{m_j} (1 - \nu_{j})^{k_j - 2 - m_j} \big]
\end{equation}
and  $\dot{B}_{k_{i}, m_{i}}(\nu_{i})$ is written similarly (by making the changes $j \to i$ and $k_j - 1 \to k_i$). Then, for $j = i$ we analyse terms in the sum over $\textbf{m}$ at the fixed point $\boldsymbol{\nu}^* = \boldsymbol{0}$: For $m_j = 0$ we have $\dot{B}_{k_j - 1, 0} (0) = 1 - k_j$, but since ${B}_{k_{i} ,m_{i}}(0) = \delta_{m_i,0}$ and $f(\textbf{k}, \boldsymbol{0}) = \boldsymbol{0}$ for $\phi > 0$  (for all threshold rules), the associated term in \eref{eq:derivGfactor} is zero. For $m_j = 1$ we have $\dot{B}_{k_j - 1, 1} (0) = k_j - 1$. Finally, for $m_j > 1$ we get $\dot{B}_{k_{j} - 1,m_{j}}(0) = 0$, so the only non-zero term corresponds to $m_j = 1$. By a similar argument, for $j \neq i$ the only surviving term in \eref{eq:derivGfactor} is $\dot{B}_{k_i, 1} (0) = k_i$ (for $m_i = 1$).

Combining these results, we can write \ref{eq:jacobMat} explicitly as  
\begin{equation}
\label{eq:jacobExp}
J^*_{ji} = -\delta_{ji} + \sum_{\textbf{k}} \frac{k_j}{c_j} (k_i - \delta_{ji})  P(\textbf{k}) f(\textbf{k}, \textbf{e}_{i}),
\end{equation}
where $\textbf{e}_{i}$ is the $i$-th basis vector of dimension $2^M-1$. The Jacobian matrix $\mathbf{J^*}$ of \eref{eq:jacobExp} encodes the structure of the multiplex, namely the degree and  overlap distributions via $P(\textbf{k})$, as well as node dynamics  (and optionally edge weights) via the response function $f(\textbf{k}, \textbf{e}_{i})$, which provides the response of a node with degree vector $\textbf{k}$ to a single infected neighbour across an $i$-type edge. The eigenvalues $\lambda_j$ of $\mathbf{J^*}$ are obtained by solving the characteristic equation $\det (\mathbf{J^*} - \lambda \mathbf{1}) = 0$. Then, the cascade condition for  complex contagion over multiplex networks (in the case $p = 0$) is
\begin{equation}
\label{eq:cascCondGen}
\text{Re}(\lambda_j) > 0
\end{equation}
for at least some $j = 1, \ldots, 2^M-1$.
Even though we cannot write an algebraic formula for $\lambda_j$ when $2^M-1 > 4$, we can compute the eigenvalues numerically. We may also find an explicit expression for the cascade condition in simple cases such as an  $M = 2$ duplex network with or without overlap, as we do in the following section.

\subsection{Eigenvalues for $M = 2$ layers}
\label{ssec:M2case}

Here we analyse the simple case of a multiplex network of $M = 2$ layers, with or without overlap ($\gamma > 0$ or $\gamma = 0$, respectively), and with Poissonian degree distributions for all composite edge types. There are three edge types; two resulting from node pairs connected in exactly one layer ($j = 1, 2$), and one composite edge resulting from node pairs connected in both layers ($j = 3$). The degree, partial degree, and weight vectors are $\textbf{k} = (k_1, k_2, k_3)$, $\textbf{m} = (m_1, m_2, m_3)$, and $\textbf{w} = (w_1, w_2, w_3)$, respectively, subject to the constraints $k = k_1 + k_2 + k_3$, $m = m_1 + m_2 + m_3$, and $w_3 = w_1 + w_2$ (assuming that weights are additive over composite edges).

We consider Poissonian degree distributions for all edge types,
\begin{equation}
\label{eq:DegDistEdgeType}
P_j(k_j) = \frac{ {c_j}^{k_j} e^{-c_j} }{ k_j ! },
\end{equation}
such that the only tunable parameter is the average degree for edge type $j$, $c_j = \sum_{k_j} k_j P_j(k_j)$. The probability-generating function of \eref{eq:DegDistEdgeType} is $G_j(u) = e^{c_j (u - 1)}$, from which the probability-generating function of the aggregate degree $k = \sum_j k_j$ takes the form $G(u) = \prod_j G_j(u) = e^{ \sum_j c_j (u - 1) }$. Then, $k$ also follows a Poisson distribution $P(k) = c^k e^{-c} / k! $ with average aggregate degree $c = \sum_j c_j$. The total number of edges a node has in layer $i = 1, 2$ is the sum of its composite edges of type $i$ plus the overlap edges of type $j = 3$, i.e. $k_i + k_3$, which is also Poisson distributed. Then, the average degrees $z_i$ in layer $i = 1, 2$ and the total average degree $z = z_1 + z_2$ are given by
\begin{subequations}
\label{eq:LayerDegs}
\begin{align}
z_i &= c_i + c_3, \\
z &= c_1 + c_2 + 2c_3.
\end{align}
\end{subequations}

As stated in the main text, we implement intimacy circle theory by considering the scaling $z_2 = \delta_z z_1$ ($\delta_z \geq 1$) and $w_2 = \delta_w w_1$ ($\delta_w \leq 1$). Since layer overlap is defined as $\gamma = |E_1 \cap E_2| / |E_1|$ with $E_i$ the edge set in layer $i = 1, 2$, we may also write $\gamma = ( N c_3 / 2 ) / ( N z_1 / 2 ) = c_3 / z_1$, where $N$ is the size of the network. Assuming that all edges in layer $i = 1, 2$ have the same weight $w_i$ (and $w_3 = w_1 + w_2$), the average weight in the network is $\langle w \rangle = \sum_j c_j w_j / c$. Overall, we can choose a set of four parameters, say $z$, $\delta_z$, $\delta_w$, and $\gamma$, together with the arbitrary constraint $\langle w \rangle = 1$, and use these relations to write the remainder of the network variables as
\begin{equation}
    \begin{cases}
        z_1 = \dfrac{z}{1 + \delta_z}\\
        z_2 = \delta_z z_1,
    \end{cases}
    \hspace{4mm}
    \begin{cases}
        w_1 = \dfrac{1 + \delta_z - \gamma}{1 + \delta_z \delta_w}\\  
        w_2 = \delta_w w_1
    \end{cases}
    \hspace{2mm}\text{and}\hspace{4mm}
    \begin{cases}
        c_1 = z_1 (1 - \gamma)\\
        c_2 = z_1 (\delta_z - \gamma)\\
        c_3 = z_1 \gamma.
    \end{cases}
\end{equation}
Finally, we write the response function $f(\textbf{k},\textbf{m})$ explicitly in the case of a multiplex network of $M = 2$ layers (Table I in main text). For a single threshold $\phi$ for all nodes in the network, the weighted sum threshold rule implies $f(\textbf{k},\textbf{m}) = 1$ for
\begin{equation}\label{eq:RespWeightedMult}
    \textbf{m} \cdot \textbf{w} \geq \phi \textbf{k} \cdot \textbf{w} \hspace{4mm} \text{and} \hspace{4mm} k > 0,
\end{equation}
and $f(\textbf{k},\textbf{m}) = 0$ otherwise. If a threshold $\phi_i$ is defined in layer $i = 1, 2$, the multiplex \textit{and} and \textit{or} rules imply $f(\textbf{k},\textbf{m}) = 1$ for
\begin{equation}\label{eq:RespMultOrAnd}
    m_{i} + m_{3} \geq \phi_{i} (k_{i} + k_{3})\hspace{4mm} \text{and} \hspace{4mm} k_{i} + k_{3} > 0,
\end{equation}
for either $i = 1$ or $2$, in the case of the \textit{or} rule, and $i = 1$ and $2$, in the case of the \textit{and} rule. Otherwise, $f(\textbf{k},\textbf{m}) = 0$. \esref{eq:DegDistEdgeType}{eq:RespMultOrAnd} allow us to write \eref{eq:jacobExp} explicitly and solve its characteristic equation, which we do below for the cases of non-overlapping and overlapping layers.

\subsubsection{Non-overlapping layers, $\gamma = 0$}
\label{sssec:M2caseNoOverlap}

 In the case of no overlap, $\gamma = 0$, $k_3 = m_3 = 0$ for all nodes, effectively reducing the Jacobian matrix $\mathbf{J^*}$ of \eref{eq:jacobExp} to two dimensions. Then, the characteristic equation $\lambda^2 - \lambda \text{Tr} \mathbf{J^*} + J_{11}^* J_{22}^* - J_{12}^* J_{21}^* = 0$ has solutions
\begin{equation}
\label{eq:2weightsEigen}
\lambda_{\pm} = \frac{1}{2} \left[ \text{Tr} \mathbf{J^*} \pm \sqrt{ ( J_{11}^* - J_{22}^* )^2 + 4 J_{12}^* J_{21}^* } \right],
\end{equation}
where $\text{Tr} \mathbf{J^*} = J_{11}^* + J_{22}^*$ is the trace of the Jacobian matrix. From \eref{eq:jacobExp} we have $J_{12}^*, J_{21}^* \geq 0$, so the eigenvalues in \eref{eq:2weightsEigen} are real numbers  (with the largest corresponding to the $+$ sign). We may write the cascade condition of  complex contagion in multiplex networks (for $p = \gamma = 0$) as
\begin{equation}
\label{eq:2weightsCascCond}
\text{Tr} \mathbf{J^*} + \sqrt{ ( J_{11}^* - J_{22}^* )^2 + 4 J_{12}^* J_{21}^* } > 0,
\end{equation}
an equation determining the region in $(\phi,z)$-space where infinitesimal perturbations can trigger global cascades~\cite{ruan2015kinetics,porter2016dynamical}.

\subsubsection{Overlapping layers, $\gamma > 0$}
\label{sssec:M2caseOverlap}

 In an overlapping multiplex network, $\gamma > 0$, with $M=2$ layers, the Jacobian matrix $\mathbf{J^*}$ of \eref{eq:jacobExp} is three-dimensional\footnote{A possible exception is the case of maximal overlap where $\gamma = 1$, and edges of type $j = 1$ are absent. This is due to the assumption that $|E_{1}| < |E_{2}|$. Here, the Jacobian can again be reduced to two dimensions, as was the case for $\gamma = 0$ where $j = 3$ edges were absent.}. The characteristic equation is $j_0 + j_1 \lambda + j_2 \lambda^2 + j_3 \lambda^3 = 0$, where
\begin{equation}
\label{eq:CharEqTerms}
j_0 = \det( \mathbf{J^*} ), \quad j_1 = -\frac{1}{2} \left[ \mathrm{Tr}^2( \mathbf{J^*} ) - \mathrm{Tr}( { \mathbf{J^*} }^2 ) \right], \quad j_2 = \mathrm{Tr}( \mathbf{J^*} ), \quad j_3 = -1.
\end{equation}
Instead of using the general methods of Cardano or Lagrange, we may find a trigonometric solution by making the affine transformation $\mathbf{J^*} = a \mathbf{A} + b \mathbf{1}$ for arbitrary constants $a$ and $b$. If $\mathbf{v}$ is the eigenvector of $\mathbf{J^*}$ associated with eigenvalue $\lambda$ ($\mathbf{J^*} \mathbf{v} = \lambda \mathbf{v}$), then $\mathbf{A} \mathbf{v} = c \mathbf{v}$ with $\lambda = ac + b$. By solving the eigenvalue problem for the affine transformation $\mathbf{A}$ we can find the eigenvalues of the original Jacobian. We take
\begin{equation}
\label{eq:AffineTerms}
a = \sqrt{ \frac{1}{6} \mathrm{Tr} ( \mathbf{J^*} - b \mathbf{1} )^2 }, \qquad b = \frac{1}{3} \mathrm{Tr}( \mathbf{J^*} ),
\end{equation}
for which $\mathrm{Tr}( \mathbf{A} ) = 0$ and $\mathrm{Tr}( {\mathbf{A}}^2 ) = 6$. The characteristic equation for $\mathbf{A}$ is $\det( \mathbf{A} ) + 3c - c^3 = 0$ with discriminant $\Delta = 4 - \det^2( \mathbf{A} )$, so we have three (distinct or multiple) real roots for $| \det( \mathbf{A} ) | < 2$. By making the change of variable $c = 2 \cos \alpha$ and using the trigonometric identity $\cos 3 \alpha = 4 \cos^3 \alpha - 3 \cos \alpha$, we finally write the eigenvalues $\lambda_l$ of the Jacobian matrix $\mathbf{J^*}$ as
\begin{equation}
\label{eq:OverlapEigen}
\lambda_l = 2a \cos \left[ \frac{1}{3} \arccos \left( \frac{1}{2} \det \left[ \frac{1}{a} \left( \mathbf{J^*} - b \mathbf{1} \right)  \right] \right) + \frac{2 \pi}{3} l \right] + b, \qquad l = 0, 1, 2.
\end{equation}
The cascade condition for complex contagion in multiplex networks (for $p=0$ and $\gamma > 0$) is for the leading eigenvalue in \eref{eq:OverlapEigen} to be positive, $\mathrm{max} \{ \lambda_l \} > 0$.

\subsection{Velocity field analysis}
\begin{figure*}
\centering
\hspace*{-6mm}
\includegraphics[scale=1]{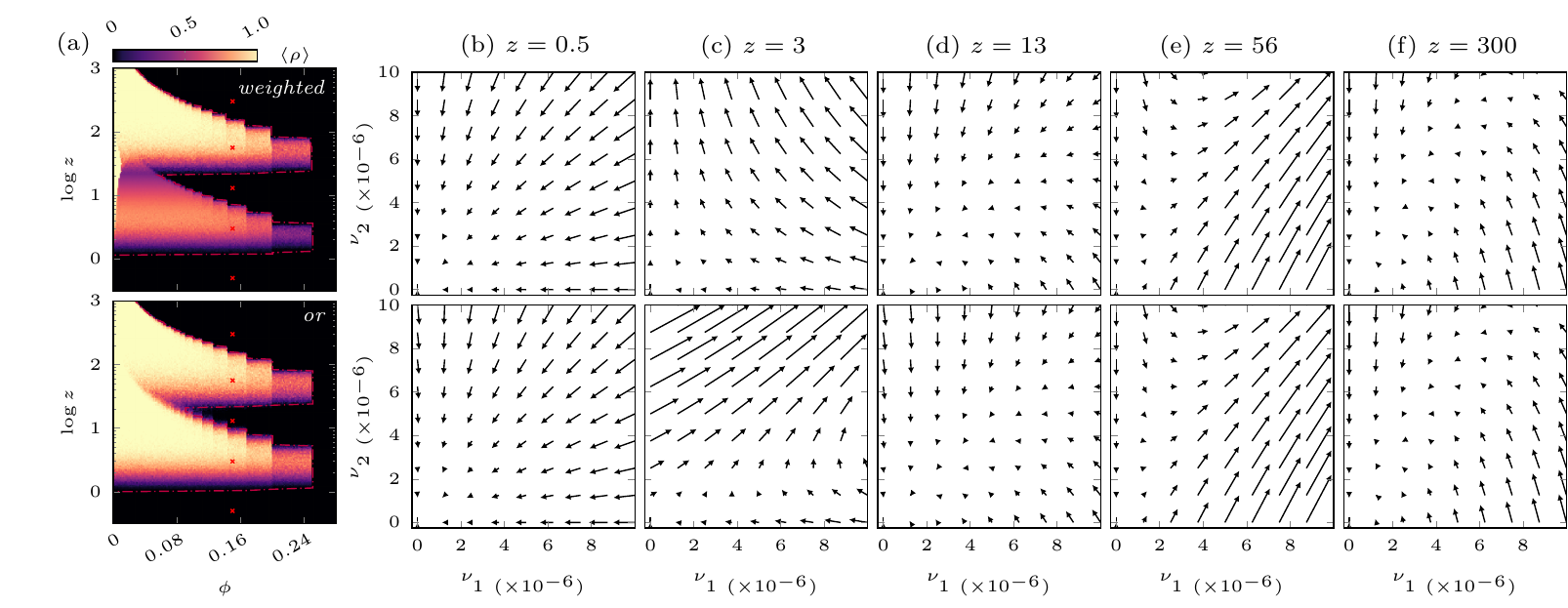}
\caption{Linear stability analysis  (a) and velocity field analysis (b--f) around the fixed point $\boldsymbol{\nu} = (\nu_1, \nu_2) = \textbf{0}$  for two threshold rules, in an $M = 2$ layer non-overlapping multiplex network with Poisson degree distributions in all layers and density skewness $\delta_{z} = 20$. Top row corresponds to the weighted sum threshold rule, and bottom row to the \textit{or} threshold rule, for otherwise identical configurations. Overlap is $\gamma = 0$, so the third component $\nu_{3} = 0$ everywhere, since composite links are absent.  The heat map in (a) shows numerical calculations of the fraction of infected nodes $\rho$ as a function of threshold $\phi$ and average total degree $z$. The dashed curve in (a)  encloses the region where \eref{eq:2weightsCascCond} is satisfied, and the velocity vectors in (b-f) are found by evaluating \eref{eq:reducedAMEs} for small $\boldsymbol{\nu}$.  In (a), numerical calculations fit analytical results perfectly.
\label{fig:multiplex_figS2}}
\end{figure*}

In this section we illustrate the typical results of the above linear stability analysis [\eref{eq:2weightsCascCond}], and compare with the output of Monte Carlo simulation, as well the velocity field of \eref{eq:reducedAMEs}, as shown in \fref{fig:multiplex_figS2}. We do this for the weighted sum threshold rule, as well as the \textit{or} threshold rule, for identical multiplexes. In \fref{fig:multiplex_figS2}(a), top and bottom, the region enclosed by dashed lines corresponds to $(\phi, z)$ configurations where the leading eigenvalue $\lambda_+$ is positive, and thus satisfies \eref{eq:2weightsCascCond}. In \fref{fig:multiplex_figS2}(b-f) we show the corresponding velocity field analysis at five points along the $\phi = 0.15$ axis: below the low-$z$ cascade phase at $z = 0.5$, within the low-$z$ phase at $z = 3$, between cascade phases at $z = 13$, within the high-$z$ phase at $z = 56$, and above the high-$z$ phase at $z = 300$. For $z = 0.5$ in both case, the system is clearly stable, with the initial condition $\boldsymbol{\nu} = \boldsymbol{0}$ being an attractor. This is due to the lack of connectivity; a giant connected component forms only at $z = 1$ for a Poisson distributed network, meaning the multiplex consists of many small, disconnected components, and a small perturbation cannot develop into a global cascade. In the lower phase, low-weight links $(i = 2)$ provide most of the connectivity, being $\delta_{z} = 10$ times more abundant, and allow for the emergence of a percolating vulnerable cluster. Hence, the system is unstable along the $\nu_{2}$ axis for both threshold rules. In the case of the weighted sum rule, the sparse but high-weight links of layer one inhibit the size of cascades driven by the sparse layer. This effect is absent for the \textit{or} rule, where layer one links only serve to facilitate cascades, resulting in the increased $\nu_{1}$ component in \fref{fig:multiplex_figS2}(c), bottom compared to top.

Between cascade regions at $z = 13$, \fref{fig:multiplex_figS2}(d), the fixed point ${\boldsymbol{\nu} = 0}$ is again an attractor, since nodes are stable to low-weight neighbour adoption from layer two, and high-weight neighbours from layer one are too sparse to percolate structurally, for both the weighted sum and the \textit{or} threshold rule. At $z = 56$, \fref{fig:multiplex_figS2}(e), nodes are mostly connected through low-weight neighbours to whom they are stable, but sparse, high-weight neighbours $(i = 1)$ now percolate structurally, and dominate the strength of adjacent nodes since weight heterogeneity is maximal in this experiment $(\delta_{w} = 10^{-3})$. As such, a percolating vulnerable cluster is able to form, and the system becomes unstable along the $\nu_{1}$ axis. Beyond this phase, at $z = 300$ for example, \fref{fig:multiplex_figS2}(f), all nodes are stable against adopting neighbours of all weights, since both layers are excessively dense.

\subsection{Comparative eigenvalue analysis}

\begin{figure}
    \centering
    \includegraphics{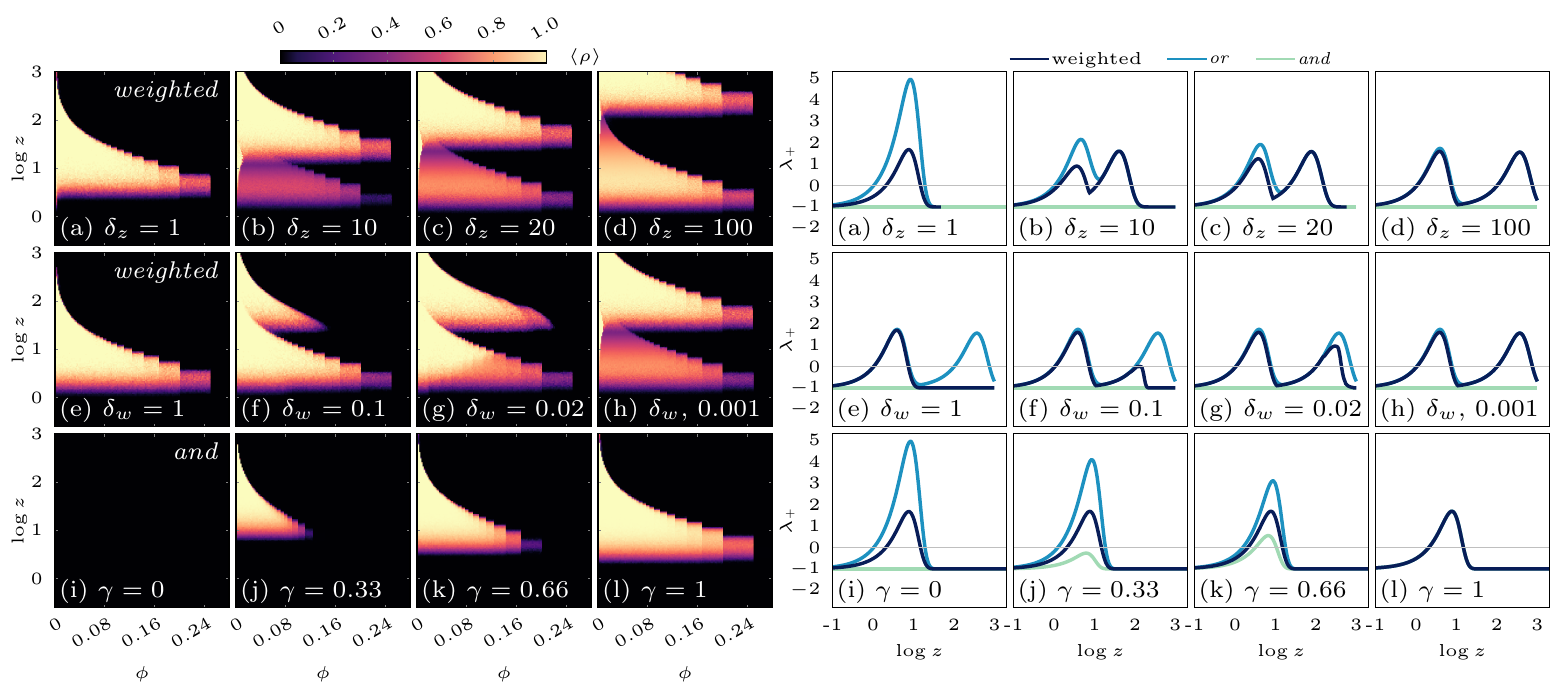}
    \caption{Comparison of Monte Carlo simulation with leading eigenvalues $\lambda_{+}$ of \eref{eq:jacobExp} as a function of increasing $\delta_{z}$, $\delta_{w}$ and $\gamma$, for each threshold rule. Heat maps for a selected threshold rule are on the left. As a general rule, $\lambda_{+}$ for the weighted sum rule is bounded above and below by the \textit{or} and \textit{and} rules, respectively. Heat maps result from $10^{3}$ single node perturbations of an $N = 10^{5}$ node multiplex. Eigenvalues are along a constant $\phi = 0.15$ slice of the corresponding heat map.}
    \label{fig:multiplex_figS14}
\end{figure}

In this section, we develop the claim that the stability of the weighted sum threshold rule is intermediate between that of the \textit{and} and \textit{or} threshold rules. We do so without proof, performing instead a comparative eigenvalue analysis of the Jacobian, \eref{eq:jacobExp}, for each of the three definitions of threshold rule, outlined in Table I of the main text. By means of elementary limiting arguments over $(\delta_{z}, \delta_{w}, \gamma)$, the space of parameters broadly defining our model multiplex, we see that the various rules converge in terms of stability at certain extremities of this space. This allows us to conclude that in general, in response to an initial perturbation our model multiplex is most stable under the \textit{and} rule, least stable under the \textit{or} rule, with the weighted sum rule providing a level of stability intermediate between the two.

In \fref{fig:multiplex_figS14}, left, we plot the Watts phase space $(\phi, z)$ along each of the axes of the parameter space $(\delta_{z}, \delta_{w}, \gamma)$. On the right of \fref{fig:multiplex_figS14}, we plot the leading eigenvalues $\lambda_{+}$ of \eref{eq:2weightsEigen} for a constant $\phi = 0.15$ slice of each Watts phase space, for each response function defined in Table I of the main text. That is, we plot $\lambda_+$ as a function of average total degree $z$ for a given $\phi$, for each threshold rule (in contrast, we do not plot Monte Carlo simulations of each threshold rule, just a representative one). When $\lambda_+ > 0$, the condition for global cascades of contagion is satisfied. Gray horizontal lines on the right of \fref{fig:multiplex_figS14} correspond to $\lambda_{+} = 0$, the value above which the system becomes unstable, and an infinitesimal perturbation triggers global cascades. When $\lambda_{+} < 0$, the system is stable, and no global cascades emerge. It is worthwhile noting that for all $\delta_{z}$, $\delta_{w}$ and $\gamma$, in the limit of $z \rightarrow 0$ and $z \rightarrow \infty$, the eigenvalues of each rule converge at $\lambda_{+} = -1$, the minimum value arising from \eref{eq:jacobExp} when the response function is $f = 0$ for all configurations $(\textbf{k}, \textbf{m})$. Trivially, this means that cascades are impossible if the multiplex is exceedingly sparse or dense, respectively.

What we observe across all values of $\delta_{z}$, $\delta_{w}$ and  $\gamma$ in \fref{fig:multiplex_figS14}, is that the \textit{and} and \textit{or} rules bound the weighted sum rule below and above, respectively, in the magnitude of the leading eigenvalue $\lambda_{+}$. In other words, the system is always least stable under an 
\textit{or} response function, and most stable under an \textit{and} response function, with the weighted sum rule intermediate between the two. In particular, these rules converge at the limiting values of $\delta_{z}$, $\delta_{w}$, and $\gamma$. Consider first the leading eigenvalue $\lambda_{+}$ under the \textit{or} rule, which is always larger than or equal to that associated with the weighted projection, as seen in the top two rows of \fref{fig:multiplex_figS14}. This can be interpreted as being due to the permissiveness of the \textit{or} rule; a node will adopt if its threshold $\phi_{i}$ is satisfied in any layer $i$. To trigger global cascades, an edge type must be of sufficient density such that it percolates structurally, but not so dense that nodes are stable against a single infected neighbour of that edge type. This condition roughly determines when the \textit{or} rules leads to global cascades. The same is true of the weighted sum rule, with the additional constraint that edges in this range of density must dominate the local neighbourhood in terms of weight. Clearly, this coupling between layers via the weighted sum rule can only serve to increase the system’s stability with respect to the or rule. This results in the \textit{or} rule being at least as unstable everywhere as the weighted sum rule. In all experiments conducted, cascading phases due to the \textit{or} rule begin earlier, and finish later as function of $z$, with respect to corresponding experiments using the weighted sum rule. This is evident for all $\lambda_{+}$, from (a) to (l) in \fref{fig:multiplex_figS14}.

It is relatively straightforward to see why eigenvalues $\lambda_{+}$ are smaller in the weighted sum rule than in the \textit{or} rule. As explained in the main text, this increased stability is due to certain ``blocked'' configurations that are formed when weight heterogeneity is large. That is, pairs of susceptible nodes connected via a high-weight edge, remainder of their neighbourhoods are week. Even if all these weakly interacting neighbours are infected, the strong interaction mutually ``immunises'' each susceptible node, ensuring that they remain in this state forever. This leads to partial cascades, evident for example in the lower phase of \fref{fig:multiplex_figS14}(b). Such cascades spread more slowly due to the presence of these immune configurations. See for example~\cite{ruan2015kinetics}, where spreading speed decreases as a result of ``blocked'' configurations. As such, the effective coupling between layers due to the weighted sum rule increases the stability of these systems with respect to \textit{or} dynamics under the same settings. In other words, the independence of each layer in the \textit{or} rule. This coupling is minimised in the $\delta_{z} \gg 1$, top row, where the effect of partial cascade diminishes [compare color of bottom phase in heat map of \fref{fig:multiplex_figS14}(b) and (f), for example], and the dynamics of the two rules converge.

Now consider the \textit{and} rule, which can be viewed as the most restrictive, requiring that a node’s threshold is satisfied in every layer before adoption takes place. When overlap $\gamma$ is zero, $\lambda_{+} = -1$ and the system is stable everywhere, illustrated in the top two rows of \fref{fig:multiplex_figS14}. This is because overlapping links, which are necessary in order for cascades to develop under the \textit{and} rule, are absent in this case. When overlap is present small perturbations can trigger global cascades, as in the bottom row where we interpolate between no overlap, and maximal overlap. Even when the \textit{and} rule allows a cascading phase, the system is more robust than the corresponding weighted sum response (with cascades emerging later and disappearing sooner in terms $z$; see third column of the bottom row in the rightmost array). An elegant illustration of the relative stability of each rule is when we $\delta_{z} = 1$, and $\gamma$ increases from $0$ to $1$. Here we observe the ``sandwiching'' of the weighted sum rule below and above by the \textit{and} and \textit{or} rules, respectively. This occurs for any value of weight heterogeneity, which is controlled by the skewness parameter $\delta_{w}$.
 
The above eigenvalue analysis highlights the contrasting stability of multiplex and aggregated systems, and even suggests that a weighted aggregate has an intermediate stability between the two multiplex behaviours, namely the \textit{and} and \textit{or} dynamics).

\section{Degree and weight distributions}

In the main text we argue that the salient features of real multiplexes, scaling in the mean degree $z_{i}$ and mean weight $w_{i}$ from layer to layer, can be modelled as  
\begin{equation}
 z_{i + 1} = \delta_{z}z_{i} \hspace{4mm} \text{and} \hspace{4mm} w_{i + 1} = \delta_{w}w_{i}.
\label{eqn:delta}
\end{equation}
We term $\delta_{z}$ the density scaling factor, or density skewness, and $\delta_{w}$ the weight scaling factor, or weight skewness. Crucially, setting $\delta_{z} > 1$ and $\delta_{w} < 1$ recovers the class of structure outlined in intimacy circle theory, indicating a multiplex growing in link density, and decreasing in mean interaction strength or weight, by layer. Both $\delta_{z}$ and $\delta_{w}$ are constant in our model, and induce exponentially distributed layer average degrees and layer average weights, since
\begin{equation}
    \lbrace z_{1},\ z_{2},\ z_{3},\ \hdots,\ z_{M} \rbrace = \lbrace z_{1},\ \delta_{z}z_{1},\ \delta_{z}^{2}z_{1},\ \hdots,\ \delta_{z}^{M - 1}z_{1} \rbrace,
\end{equation}
with a similar expression holding for the distribution of $w_{i}$ values. Although $z_{1}$ appears in the latter expression, this choice is arbitrary and it is not in fact a free parameter of our experiments. For example, we could equally have written $\lbrace \delta_{z}^{-1}z_{2},\ z_{2},\ \delta_{z}z_{2},\ \hdots,\ \delta_{z}^{M - 2}z_{2} \rbrace$. In experiments we set the total average connectivity $z$, defined as $z \equiv z_{1} + z_{2} + \hdots + z_{M}$, as well as the density skewness $\delta_{z}$. Since each expression in \eref{eqn:delta} has only two degrees of freedom, choosing $z$ and $\delta_{z}$ effectively prescribes the individual layer averages $z_{1},\ z_{2},\ \hdots,\ z_{M}$. Similarly, the distribution of weight means $w_{1}, \hdots, w_{M}$ has only two degrees of freedom. As for the distribution of $z_{i}$ values, we do not explicitly set $w_{i}$, rather, we constrain the average weight $\langle w \rangle$, which along with the weight scaling constant $\delta_{w}$, determines each $w_{i}$. The total mean weight $\langle w \rangle$ over the multiplex can be found by summing over edge type means, $\langle w \rangle = \sum_{j} c_{j}w_{j} / c$, where $c_{j}$ is the average degree of the $j$-th edge type, equal to $z_{i}$ in the case of zero overlap, and $w_{j}$ is the sum of edge weights constituting the resultant edge of type $j$. In all experiments throughout this work we set the network-wide average weight to $\langle w \rangle = 1$. This allows us to isolate the effect of varying the skewness $\delta_{z}$ in edge density, and the skewness $\delta_{w}$ in interaction strength, across layers. Note that we impose the additional constraint that all edge weights be positive.

\subsection{Maximal weight heterogeneity}

\begin{figure}
	\centering
	\includegraphics[scale=1]{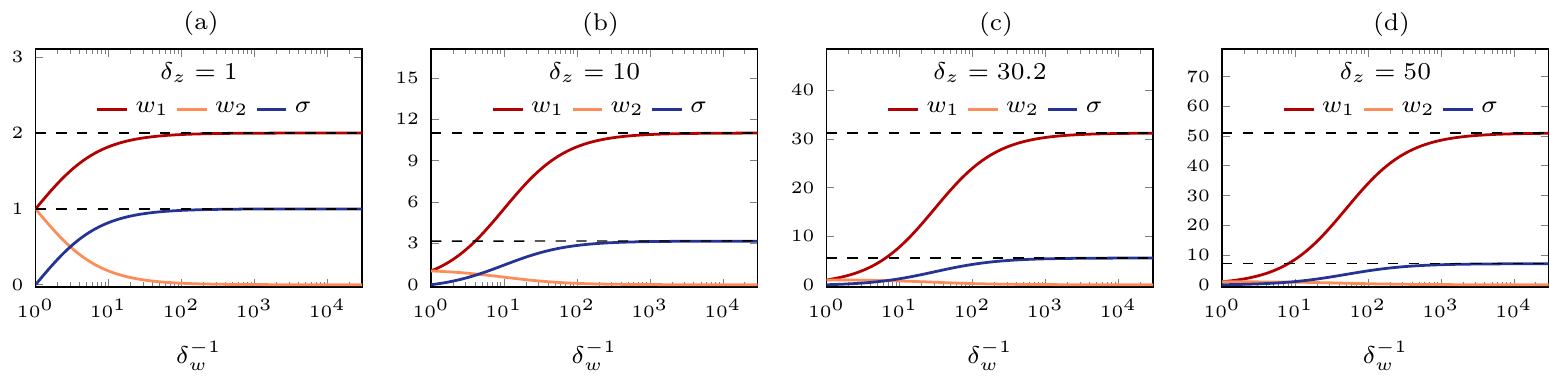}
	\caption{The dependence of weight means $w_{i}$ and standard deviation $\sigma$ of the weight means on $\delta_{w}$. $M = 2$ layers. 
\label{fig:multiplex_figS5}}
\end{figure}
In this section we discuss the consequences of our definition of average degree and weight scaling. In particular, given the constraints $\langle w \rangle = 1$ and $w_{i} > 0$ $\forall i$, the weight scaling constant $\delta_{w}$ is defined over the interval $(0, 1]$, with $\delta_{w} = 1$ giving identical mean weights $w_{i}$ in each layer, and $\delta_{w} \rightarrow 0$ tending to maximal weight heterogeneity. The interpolation between these limits is shown in \fref{fig:multiplex_figS5} for the weight means $w_{i}$ and the standard deviation of the means\footnote{Not to be confused with $\sigma_{w_{j}}$, the weight standard deviation of edge type $j$} $\sigma$, for various values of $\delta_{z}$ used in this work. In $M = 2$ layers, assuming density skewness $\delta_{z}$, it is straightforward to show that the weight heterogeneous limit of these quantities is
\begin{eqnarray}
    \lim_{\delta_{w} \rightarrow 0}w_{1} &=& 1 + \delta_{z},\\
    \lim_{\delta_{w} \rightarrow 0}w_{2} &=& 0,\\
     \lim_{\delta_{w} \rightarrow 0}\sigma &=& \sqrt{\delta_{z}},
\end{eqnarray}
where the standard deviation of edge weights across the multiplex is defined as $\sigma^{2} = \sum z_{i}\left(w_{i} - \langle w \rangle \right)^{2}/z$. Clearly, increasing weight heterogeneity by decreasing $\delta_{w}$ has a saturating effect on the values $w_{i}$ and $\sigma$. Furthermore, decreasing $\delta_{w}$ below the limiting value will have diminishing effect on the actual threshold processes evolving over the network. In the experiments described in the main text, we first set a density skewness $\delta_{z}$, and then increase weight skewness subject to the constraints that all weight values are positive ($w_{i} > 0$ for all layers $i$) and that $\langle w \rangle = 1$. In Fig. 2 of the main text, for example, we are interested in the approach towards maximal weight heterogeneity from a uniform distribution given by $\delta_{w} = 1$. The maximal weight distribution leads to results given by the white contour in Fig. 2(c), corresponding to a value of $\delta_{w} = 10^{-6}$. Again, any positive value in the range $0 < \delta_{w} < 10^{-6}$ would give identical results due to the limiting effect of $\delta_{w} \rightarrow 0$. In \fref{fig:multiplex_figS9} of the present text, we conduct a similar experiment in an empirical Twitter network. There, we study the approach to maximal weight heterogeneity $\delta_{w} \rightarrow 0$, beginning with a uniform weight distribution $\delta_{w} = 1$. 

\subsection{Weak and strong conditions for reentrant phases}

\begin{figure}
    \centering
    \includegraphics[scale=1]{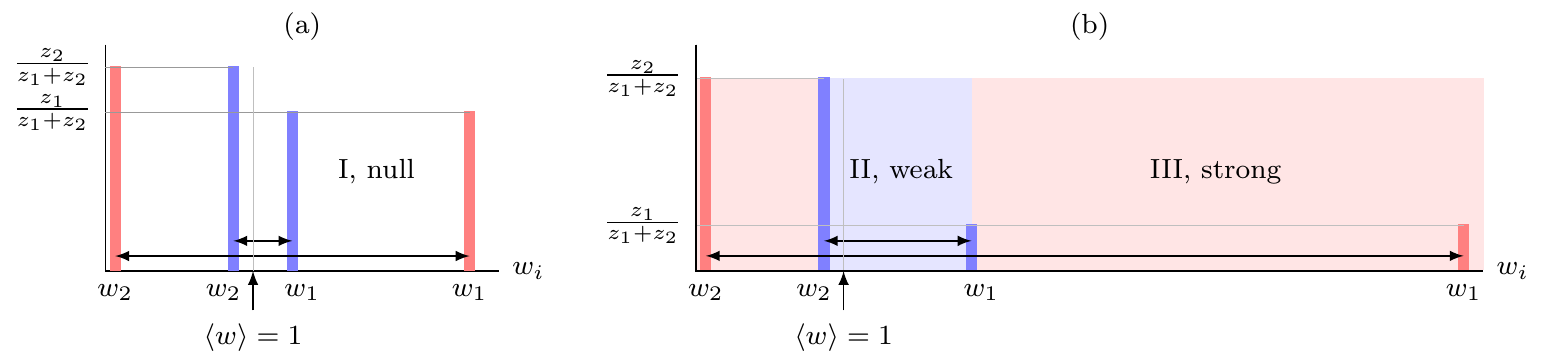}
    \caption{Conditions for the observations of a cascading phase at high $z$, resulting in reentrant phase transitions. The null case, when $\delta_{z} \sim 1$, does not exhibit reentrant transitions, under any dynamic. If the weak condition is met, $\delta_{z} > 1$ but with low weight variance $\delta_{w} \sim 1$, reentrant transitions only emerge under the multiplex-\textit{or} case, where there is the increased cost of considering thresholds in individual layers. When the strong condition is met, $\delta_{z} > 1$ and $\delta_{w} < 1$, reentrant transitions occur under all dynamics.}
    \label{fig:multiplex_figS8}
\end{figure}

In this section we summarise the conditions on $\delta_{z}$ and $\delta_{w}$ such that a cascading regime at high $z$ is formed, resulting in reentrant phase transitions in cascade size for constant $\phi$ intervals in $(\phi, z)$ space. By way of illustration, we consider an $M = 2$ layer multiplex and vary $\delta_{z}$ and $\delta_{w}$, demonstrating three behavioural regimes as follows. Case I represents the null condition in \fref{fig:multiplex_figS8}(a), where reentrant transitions are not observed in either the weighted sum threshold rule, or the \textit{or} threshold rule. In case II, reentrant transitions are observed only under multiplex-\textit{or} dynamics, called the weak condition in \fref{fig:multiplex_figS8}(b). In case III, reentrant transitions are observed under both the weighted and multiplex-\textit{or} dynamics, called the strong condition in \fref{fig:multiplex_figS8}(b). We show the values $w_{i}$ and their masses $z_{i} / (z_{1} + z_{2})$, for a low-$\delta_{z}$ configuration in (a), and a high $\delta_{z}$ configuration in (b), each comparing a low and high variance weight distribution (blue and red masses, respectively). For the purposes of this illustration we can assume layer overlap $\gamma = 0$. Further, we use synonymously the expressions ``high-$z$ cascades'' and reentrant phase transitions.

If density scaling $\delta_{z}$ is not substantial, as is the case in \fref{fig:multiplex_figS8}(a), then no reentrant transitions occur regardless of the adoption rule. Consider first the dynamics of the weighted sum threshold rule, in a system where all nodes are in the susceptible state except for an infinitesimal seed. If $\delta_{z}$ is close to unity, a node $u$ with a single infected neighbour of the strong type is unlikely adopt. That is because when $\delta_{z} \sim 1$, $u$ is likely to have other strong neighbours contributing to its overall influence, meaning one infected neighbour is insufficient to overcome the threshold $\phi$, even if that neighbour is of the strong type. This is the case even when weight heterogeneity is maximal, i.e., when $\delta_{w} \rightarrow 0$. Likewise, in the dynamics of the multiplex \textit{or} rule when $\delta_{z} \sim 1$, a node is likely to have a similar number of neighbours of each type. At high-$z$, this means a typical node configuration has large number of links both layers, meaning that a small perturbation is unlikely to satisfy the threshold in any layer. We thus term $\delta_{z} \sim 1$ the null condition.

Now consider the case where $\delta_{z}$ is substantially larger than $1$, but where weight heterogeneity between layers is low, $\delta_{w} \sim 1$. This corresponds to the weak condition, case II in \fref{fig:multiplex_figS8}(b), where reentrant transitions are observed in the $or$ rule, but not in the weighted sum rule. Again, let us first consider the weighted sum threshold rule in the presence of an infinitesimal seed. At high-$z$ configurations, a node is likely to have a small number of links from the strong layer, and a large number of links from the weak layer. If a node $u$ has only one neighbour from the strong layer, and that neighbour is infected, its influence may still be insufficient to overcome the threshold $\phi$ and infect $u$, as long as the condition $\delta_{w} \sim 1$ is in place. That is, the sum of the influence from the weak neighbours of $u$, which are numerous when $\delta_{z} > 1$, is enough for $u$ to remain stable. In contrast, the \textit{or} rule readily leads to infection of $u$ in this setting. By considering its configuration layer by layer and applying the threshold rule, $u$ adopts due to all influence in the strong layer being infected. However, this comes with a trade-off in complexity; $u$ now has to consciously consider $M$ layer configurations, and make $M$ decisions.  

Finally, consider case III in \fref{fig:multiplex_figS8}(b), identical to that described above, except that weight heterogeneity between layers is now substantial, or $\delta_{w} < 1$. Under the weighted sum threshold rule, the scenario described above may now lead to the infection of node $u$. This is because the infected influence of the one strong neighbour now overwhelms the influence of the susceptible weak neighbours, despite them being more numerous. Cascades still occur in the \textit{or} case, since this effect was driven by the density scaling factor $\delta_{z}$, which has not changed. As such, we recover the strong condition for the observation of reentrant phase transitions. Namely that when  $\delta_{z} > 1$ \textit{simultaneously} with $\delta_{w} < 1$, the system is vulnerable to cascades at high-$z$ regardless of the functional form of the threshold rule. This leads to reentrant transitions, where the high-$z$ cascading regime is separated from the low-$z$ cascading regime by an intermediate stable phase. As shown in Fig. 4(a) and (b) in the main text, the actual size of this intermediate phase depends on the magnitude of $\delta_{z}$.

We conclude that a necessary condition for the observation of high-$z$ cascades is that $\delta_{z} > 1$, regardless of whether dynamics are defined by a weighted sum threshold rule, or the \textit{or} multiplex rule. We refer to this as the weak condition, since it is necessary for high-$z$ cascades under all response functions, but sufficient only in the case of the \textit{or} threshold rule. When the weak condition is not met, we recover the null case of \fref{fig:multiplex_figS8}(a). In node dynamics following the weighted sum threshold rule, a necessary and sufficient condition for the emergence of high-$z$ cascades entailing reentrant phase transitions is that $\delta_{z} > 1$ simultaneously with $\delta_{w}  < 1$. Of course, if these conditions are satisfied, dynamics following the \textit{or} threshold rule also lead to reentrant phases.

\subsection{Poisson and log-normal degree distributions}

The scaling conditions in \eref{eqn:delta} specify only the average $c_{j}$ of the degree distribution of edge type $j$, so we are free to choose the actual form of the distribution. Specifically, we use Poisson distributions (PO) in Fig. 4 of the manuscript, as well as Figs. S\ref{fig:multiplex_figS2} and S\ref{fig:multiplex_figS14} of the present text. Otherwise, we use log-normal distributions (LN). A Poisson degree distribution is prescribed entirely by its mean $c_{j}$, 
\begin{equation}
    P_{j}(k_{j} \mid c_{j}) = e^{-c_{j}} \dfrac{c_{j}^{k}}{k_{j}!},
\end{equation}
which is defined for integer $k_{j}$, as required, and the average degree $c_{j}$ of each edge type $j$. In contrast, the log-normal distribution is defined for continuous variables, so it remains to define an appropriate discretisation. Since the structural percolation transition is of critical importance to the Watts model and its extensions, any discretisation must allow for a non-zero mass of $k = 0$ degree nodes (which is undefined in the continuous distribution). This rules out, for example, discretisations based on the ceiling function, since values in the range $0 < k < 1$ will be rounded to $k = 1$, resulting in $P_{i}(0) = 0$.  With that in mind, we generate discrete $k$ by rounding to the nearest integer. Other choices of discretisation are of course possible. In the numerical construction of a configuration model network with LN degree distribution, it is straightforward to randomly sample continuous $k$, and round up or down to the nearest integer. The masses of the $k$ values thus obtained can be found by integrating the log-normal probability density function over the interval $(k - \tfrac{1}{2}, k + \tfrac{1}{2})$, or equivalently, evaluating the cumulative distribution at the limits. As such, degrees $k_{j}$ are distributed as
\begin{equation}
    P_{j}(k_{j} \mid c_{j}, \sigma_{k_{j}}) = F(k_{j} + \tfrac{1}{2} \mid c_{j}, \sigma_{k_{j}}) - F(k_{j} - \tfrac{1}{2} \mid c_{j}, \sigma_{k_{j}}),
\end{equation}
where $F$ is its cumulative distribution of the continuous log-normal distribution, which we assume is defined $F(k_{j}) = 0$ for $k_{j} \leq 0$. The actual value of $F$ is given by the error function
\begin{equation}
    F(k_{j} \mid c_{j}, \sigma_{k_{j}}) = \dfrac{1}{2} + \dfrac{1}{2}\text{erf}\left( \dfrac{\ln k_{j} - c^{\prime}_{j}}{\sqrt{2}\sigma^{\prime}_{k_{j}}} \right),
\end{equation}
where $c^{\prime}_{j}$ and $\sigma^{\prime}_{k_{j}}$ are the mean and standard deviation of the underlying normal distribution\footnote{Normal and log-normal moments are related through $\mu^{\prime} = \log \left[ \mu \left(1 + \tfrac{\sigma^{2}}{\mu^{2}}\right)^{-\tfrac{1}{2}} \right]$ and $\sigma^{\prime} = \left(1 + \tfrac{\sigma^{2}}{\mu^{2}}\right)^{\tfrac{1}{2}}$.}. This approach leads to precise agreement between the simulated multiplex and analytic solution, and most importantly, allows for a structural percolation transition in the resulting configuration model multiplex, at low $z$ due to the presence of $k_{j} = 0$ degree nodes.

Finally, since our experiments generate log-normal distributions with means $c_{j}$ spanning several orders of magnitude, we must select experimental values of standard deviation $\sigma_{k_{j}}$ appropriately. That is, the standard deviation must scale with the average, and we set $\sigma_{k_{j}} = 2c_{j}$. To see why this is necessary, consider that $\sigma_{k_{j}} = 10$ is a relatively high variance if the average degree $c_{j} = 1$, with a qualitatively broad spread. However, if $c_{j} = 1000$, a standard deviation of $\sigma_{k_{j}} = 10$ leads to a very narrow overall distribution, qualitatively giving a sharp peak rather than a broad tail.

\subsection{Uniform and log-normal weight distributions}
\begin{figure}
    \centering
    \includegraphics{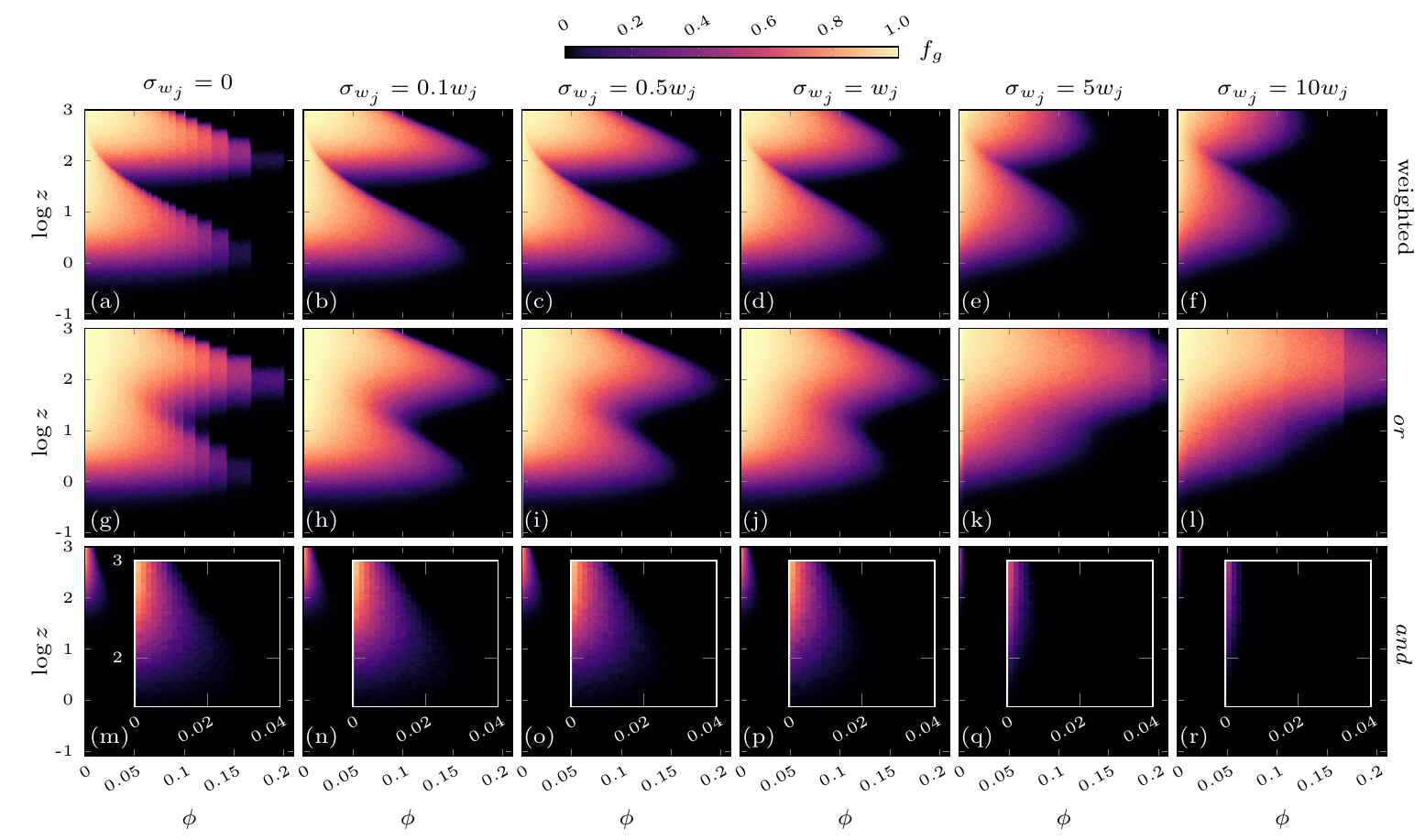}
    \caption{Varying the standard deviation of weights in layer $i$, using a log-normal weight distribution with average $w_{i}$ and standard deviation $\sigma_{w_{j}}$. The first column corresponds to Fig. 3(a-c) of the main text, with $\sigma_{w_{j}} = 0$ recovering the uniform weight distribution. We use the weighted sum threshold rule in row one, the \textit{or} threshold rule in row two, and the \textit{and} threshold rule in row three. Networks are of size $N = 10^{5}$, and we record the frequency $f_{g}$ of global cascades after $10^{3}$ realisations of single node perturbation.
    \label{fig:multiplex_figS13}}
\end{figure}

In the main text, for simplicity of presentation as well as ease of analytic solution, we assume that the weight distribution with each layer $i$ is uniform with value $w_{i}$. In this section we verify the robustness of our principal results when the weight distributions on each layer are no longer uniform. In \fref{fig:multiplex_figS13}, we progressively increase the standard deviation $\sigma_{w_{j}}$ of a log-normal weight distribution \textit{within} each layer with mean $w_{j}$, with row one using the weighted mutliplex rule, row two using the multiplex \textit{or} rule, and row three using the multiplex \textit{and} rule. The mean weight $w_{j}$ in layer $j$ is determined as before. That is, the means of the weight distributions within each layer are related by $w_{i + 1} = \delta_{w}w_{i}$. Since weights in layer $i$ have mean $w_{i}$, and we have applied the constraint $\langle w \rangle = 1$, the system wide average is also $1$. The means between layers can be made more heterogeneous by decreasing $\delta_{w} \in (0,1]$, and the distributions within layers made more heterogeneous by increasing $\sigma_{w_{j}} \in [0,\infty)$. As is the case with log-normal degree distributions, we provide a scale to the weight standard deviation by varying it with respect to the mean $w_{j}$. 

When $\sigma_{w_{j}} = 0$, we recover Figs. 3(a-c) in the main text, where edge weights were uniform with layers. This provides the first column in \fref{fig:multiplex_figS13}, namely plots (a), (g) and (m). As discussed in the main text, when node dynamics are determined by the \textit{and} and \textit{or} rules, i.e., the second and third rows of column one, the multiplex is effectively unweighted. In this case, high-$z$ cascades, and the associated reentrant phase transition, are entirely driven by layer density skewness $\delta_{z}$ along with individual layer thresholds.

Increasing $\sigma_{w_{j}} = 0$ across the top row of \fref{fig:multiplex_figS13}(a-f), using the weighted sum rule, reentrant transitions are observed to emerge even in the presence of very large weight heterogeneity within layers. This is surprising, and highlights that it is the first moment of the layer weight distributions, the mean $w_{i}$, along with the inter-layer skewness $\delta_{w}$ that drive high-$z$ cascades under this threshold rule. Similarly, reentrant phase transitions persist under the \textit{or} threshold rule, as seen along row two of \fref{fig:multiplex_figS13}(g-l), although are not visible under the most strongly heterogeneous distributions ($\sigma_{w_{j}} = 5 w_{j}$ and $10w_{j}$). Further, we observe that the cascading phase shrinks to lower and lower $\phi$, for increasing $\sigma_{w_{j}}$ in row one. In contrast, the effect of weight heterogeneity in the \textit{or} rule in row two is to extend the cascading phase to higher $\phi$. This is surprising, and represents another instance where explicit multiplexity leads to global cascades in settings where a pure weighted structure cannot.

Finally, in row three of \fref{fig:multiplex_figS13}(m-r), where node dynamics are determined by the \textit{and} rule, we see that the effect of intra-layer weight heterogeneity is to diminish the size of the cascading phase. That is, for larger and larger $\sigma_{w_{j}}$, the cascading phase extends to smaller and smaller $\phi$. This was the case in row one, for the weighted sum rule, and the opposite of the case in row two for the \textit{or} rule. 

\section{Monte Carlo simulation of binary-state dynamics}
\begin{figure}
    \centering
    \includegraphics{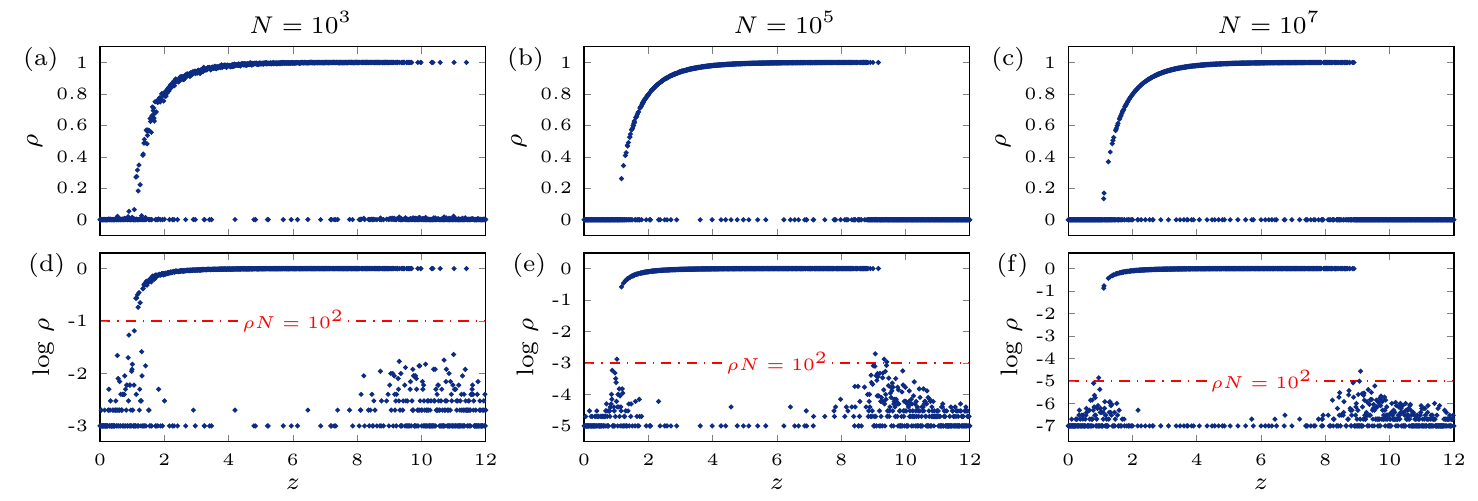}
    \caption{Individual realisations of single node perturbation on an $M = 2$ layer multiplex with $\delta_{z} = 1$, $\delta_{w} = 1$, overlap $\gamma = 0$, with Poisson degree distributions on each layer. Vertical plots illustrate identical data, linear and log scale. Horizontal dashed lines show a low-$\rho$ region of constant size or ``local'' cascades, where cascade size $\rho \sim 10^{2} / N$ does not scale with $N$.}
    \label{fig:multiplex_figS11}
\end{figure}

We implement numerically a multiplex network using the multivariate configuration model, which entails $2^{M} - 1$ independent applications of the single layer configuration model, one for each composite link type. Since layer overlap is accounted for by the use of composite edges, we do not allow for double edges in the resultant network. Complex contagion is implemented numerically via Monte Carlo simulations of a monotone binary-state dynamics, where nodes are selected uniformly at random for update in asynchronous order, generating a series of time steps. Once a node state changes from susceptible to infected, it remains so for the rest of the dynamics, thus ensuring a steady state in a finite simulation. Each time step consists of $N$ node updates, where a randomly selected node adopts only if the threshold rule is satisfied.

In all experiments in this paper, we are interested in the steady state of the system after single node perturbation. This state is captured by $\rho$, the total fraction of the $N$ nodes in network that are infected at $t \rightarrow \infty$. In \cite{watts2002simple}, Watts defines a cascade to be ``global'' if its size is non vanishing in the infinite network limit. In other words, the cascade size is not constant, and occupies a positive fraction of an infinite network. It is straightforward to identify local cascading regimes in \fref{fig:multiplex_figS11} via finite size scaling. Here, we plot the outcome $\rho$ of individual realisations, over a range of $z$ and constant $\phi$. Clearly, the noise at $\rho \rightarrow 0$ is of constant size, occupying a range $0 < \rho N < 10^{2}$, regardless of network size. A simple approach to distinguishing local and global cascades is to simulate sufficiently large networks (we choose $N = 10^{5}$, $10^{6}$ and $10^{7}$ throughout this work), and to set a relatively high threshold $\rho_{g}$ for what constitutes a global cascade. In (b,c) and (e,f), given the magnitude of $N$, cascades larger than $\rho_{g} = 10^{-2}$ will with high probability be global, and scale with network size. This allows us to define $f_{g}$, the frequency of occurrence of global cascades, used throughout this work. We also use $\langle \rho \rangle$, the expected final size of cascades that are determined global by the cutoff $\rho_{g}$.

Finally, it is useful to note that global cascades, once they occur, display very little variance in size. This can be seen in \fref{fig:multiplex_figS11}(a-c), with variance of global cascade size diminishing for larger and larger $N$. As such, error bars indicating the variance after multiple realisations would be smaller than the point sizes in (b) and (c).

\section{Twitter network}
\begin{table}
    \centering
	\caption{Number of occurrences and average degree of each multiplex edge type in an empirical Twitter multiplex. $E_{1}$ and  $E_{2}$ are the sets of edges in the mutual-mention and follower networks, respectively. Sizes of $E_{10} \equiv E_{1} \setminus E_{2}$, $E_{01} \equiv E_{2} \setminus E_{1}$ and $E_{11} \equiv E_{1} \cap E_{2}$ are shown, and $N = 370,544$.\label{tab:tab1}}
		\begin{tabular}{c c c c c c c}
		\hline\hline
	    & $E$ & $E_{1}$ & $E_{2}$ & $E_{10}$ & $E_{01}$ & $E_{11}$\\
	  	\hline 
	  	$|\bullet|$ & $30,717,559$ & $999,182$ & $30,168,645$ & $548,914$ & $29,718,377$ & $450,268$\\
	    $z$ & $165.8$ & $5.393$ & $162.8$ & $2.963$ & $160.4$ & $2.430$ \\
		\hline\hline
		\end{tabular}
\end{table}
We validate out model on an empirical Twitter dataset, which was collected over the period of June 2014 and October 2018 through the Twitter Powertrack API provided by Datasift with an access rate of 15\%. The data records microblog posts of $140$ characters, called tweets, posted in French in the GMT and GMT+1 time zones, together with user profile information. In order to construct a multiplex representation of the proxy social network, first we followed user interactions defined as direct mentions. In Twitter, a mention represents a direct interaction between users in the content of a tweet using the @ symbol (@username). When a user $u$ mentions another user $v$, the tweet containing the mention is visible directly in the feed of user $v$. After creating the network of users who at least once mentioned each other mutually during the observation period, we extracted the second largest connected component of this structure for further investigation. The obtained network contained $N = 370,544$ nodes and $999,182$ mutual-mention links. In order to construct a multiplex structure, we considered as a second layer of interaction all follower/followee links between the same set of users, which in turn provided us $30,717,559$ links. We argue that while the first mutual-mention layer corresponds to the relatively sparse but strongly interacting layer, the second layer corresponds to the densely connected but weakly interacting layer in our model.\footnote{Following the data handling policy of the company and the GDPR regulations of the EC regarding privacy, the utilized dataset cannot be shared directly. However, similar dataset can be collected via the open API maintained by Twitter or could be constrained using already open datasets.}
\begin{figure}
    \centering
    \includegraphics[scale=1]{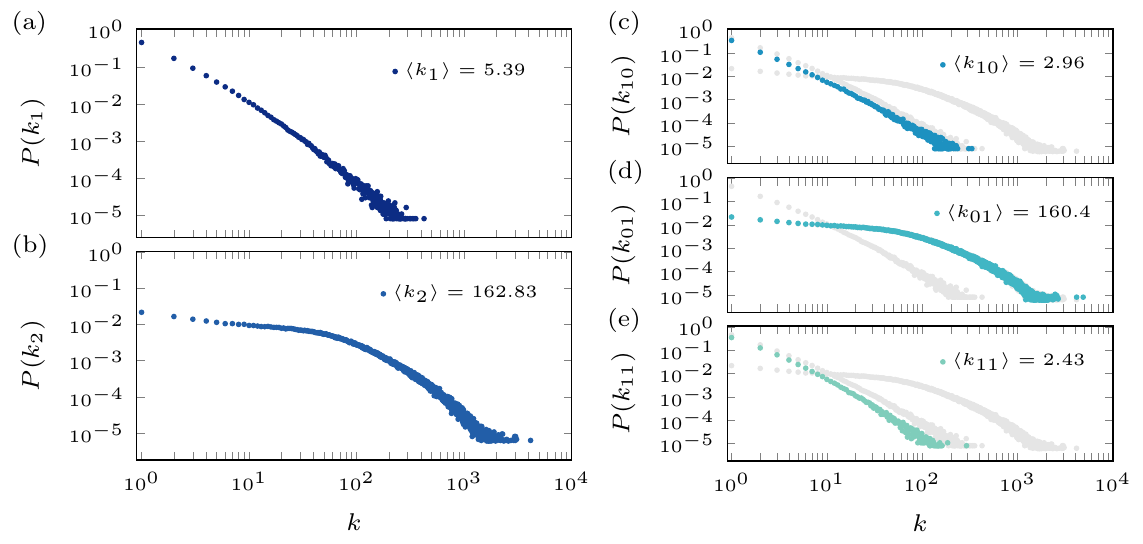}
    \caption{Degree distributions for an empirical Twitter dataset. Plots (a) and (b) show layer edge types $i = 1$ and $2$, (c) to (e) the resultant edge types $j = 1$, $2$ and $3$ due to layer overlap.}
    \label{fig:multiplex_figS6}
\end{figure}

As such, the Twitter multiplex qualitatively supports intimacy circle theory. Consider that mutual-mentions consume both the time and cognitive capacities of each user involved, constraining the number of mentions that can be made by a given individual. In contrast, following another user is inexpensive in terms of time and concentration, and a greater quantity of these relationships can be afforded. Once a user follows another, the cost of maintaining that link is minimal. It can be maintained passively, unlike mutual-mention relationships. As expected, the average mutual-mention degree of $\langle k_{1} \rangle = 5.39$ is much smaller than the average follower degree of $\langle k_{2} \rangle = 162.83$. In our model, the degree skewness factor therefore equals $\delta_{z} = 30.2$. 

The various degree distributions for the Twitter multiplex are shown in \fref{fig:multiplex_figS6}. The degree distributions within layers are shown in \fref{fig:multiplex_figS6}(a) and (b), and are clearly broad tailed. Given that the densities in each layer are relatively skewed, with $\delta_{z} = 30.2$, and that edge set overlap is substantial, with $\gamma = 0.45$, the degree distributions of the three resultant link types are well approximated by the degree distributions within each layer, \fref{fig:multiplex_figS6}(c-e). That is, the distributions of $k_{10}$ and $k_{01}$ degrees follow closely those of layer one and layer two degrees, respectively. Finally, the degree $k_{11}$ of the overlapping edge type is well approximated by that of the sparse layer, layer one. This is to be expected given the density skewness $\delta_{z} = 30.2$, since overlapping links constitute a much larger sample of layer one than layer two links, proportionally. In other words, $|E_{11}| / |E_{1}| \simeq 0.45$ whereas $|E_{11}| / |E_{2}| \simeq 0.015$. The former quantity provides the overlap $\gamma = 0.45$.

\subsection{Sparsification}

Throughout the main text, we are interested in studying the dependence of the global cascade condition on the average degree $z \equiv z_{1} + z_{2}$ of the underlying multiplex. That is, we wish to explore a phase space $(\phi, z)$, which requires producing networks of a desired average degree $z$. The Twitter network as collected has an average degree of $165.8$, and thus requires sparsification to produce samples with a desired average degree lower than this initial value, and densification for average degrees that are higher. Any choice of algorithm can only approximate the network at higher and lower $z$, since we do not have access to historical data showing the multiplex at differing levels of connectivity, nor do we know how the network will evolve beyond its actual state. As such, sparsification and densification algorithms must be used to suggest extrapolations of the empirical network to desired values of connectivity $z$, with the caveat that each algorithm introduces its own biases.

To obtain average degrees of $0.1 \leq z < 165.8$, we sparsify by removing links uniformly at random. Quite simply, we randomly select and remove links sequentially until the original empirical network is reduced to the desired $z$ value. This has the advantage that certain correlations are preserved, such as degree-degree correlations, clustering and community structure. Importantly, this algorithm preserves density skewness $\delta_{z}$ and overlap $\gamma$, while keeping the overall shape of the degree distribution relatively unaltered. As shown in Fig. 3(d-f) of the main text, sparsified Twitter multiplexes behave in accordance with predictions made on configuration model networks. That is, using the weighted sum and the \textit{or} threshold rules, the sparsifying network undergoes three transitions; the first by exiting the upper cascading phase, the second and third by entering and exiting the lower cascading phase, respectively. Using the \textit{and} multiplex rule the sparsified network passes through a single transition by exiting the cascading phase [Fig. 3(f)], as expected given the Twitter network's layer density skewness $\delta_{z} = 30.2$ and overlap $\gamma = 0.45$

\subsection{Densification}

Preliminary experiments indicate that at its original value of $z = 165.8$, the multiplex is susceptible to global cascades under the weighted sum rule, as well as the \textit{or} multiplex rule, over a large interval of thresholds $\phi$. In particular, since both layers have average degree $z_{i} > 1$, we expect the observed network to be situated within the upper cascading phase. By applying the sparsification algorithm of the previous section, this is indeed found to be true; the network undergoes three phase transitions in susceptibility to global cascades when lowering $z$ from $165.8$, to $0.1$ [see below the horizontal dashed line in \fref{fig:multiplex_figS9}(f), and Fig. 3(d-f) of the main text]. The goal of this section will be to show that by increasing the connectivity of the multiplex, the fourth and final phase transition is traversed. That is, by increasing its average degree, the empirical multiplex can be made stable against global cascades, thus exiting the upper cascading phase.
\begin{algorithm}[]
\DontPrintSemicolon
\SetKwInOut{Input}{input}\SetKwInOut{Output}{output}
    $G \leftarrow G_{0}$\;
\While{$z_{G} < z$}{
    $v \leftarrow$ node chosen u.a.r from $V$\;
    $V_{s} \leftarrow $ Burn$(G, v)$\;
    $E_{1} \leftarrow E_{1} \cup \lbrace (v, w) \mid (w, i, j) \in V_{s} \text{ and } i = 1 \rbrace$\;
    $E_{2} \leftarrow E_{2} \cup \lbrace (v, w) \mid (w, i, j) \in V_{s} \text{ and } j = 1 \rbrace$\;
    $G \leftarrow (V, E_{1}, E_{2})$\;
    $z_{G} \leftarrow$ average degree of $G$\;
}
\caption{Forest-Fire Process$(G_{0} = (V, E_{1}, E_{2}), z)$ \label{alg:alg1}}
\end{algorithm}

To extrapolate the empirical Twitter network to average degrees $165.8 < z \leq 1000$, it is not desirable to add links uniformly at random. This is because by the time it grows to $z = 1000$, the network will be almost entirely random, significantly reducing the correlations typical of empirical networks. To incorporate the original structure and preserve as much as possible empirical correlations, we use a model of densification known as the forest-fire process~\cite{leskovec2007graph,kanade2016distance}, which is described in Algs. \ref{alg:alg1} and \ref{alg:alg2}. This process amounts to an extrapolation of the original network to higher $z$, through the probabilistic addition of links biased by the existing structure. The forest-fire model originally proposed by Leskovec \textit{et al.} in \cite{leskovec2007graph} has a simple intuitive justification. It is based on having new nodes attach to the network by ``burning'' through existing edges in an epidemic manner. For example, a new node $u$ attaches to a randomly selected node $v$ in the existing graph, and begins burning through the edges of $v$, attaching to any new node it encounters following a certain probability distribution. In the context of Twitter network growth, this would be interpreted as a new user randomly selecting initial accounts to follow, then browsing the followers of those users in order to find additional accounts to follow, which he does with some probability. The new user continues recursively, using accounts discovered in previous steps to extend their list of contacts, until the process dies out.

\begin{algorithm}
\SetKwInOut{Input}{input}\SetKwInOut{Output}{output}
\Output{set $V_{s}$ of stubs $(u, i, j)$, tuple of $u \in V$ and $i,j \in \lbrace 0, 1 \rbrace$ indicating $(u, v) \in E_{1}$, $E_{2}$}
 $V_{s} \leftarrow \emptyset$\;
 $D \leftarrow \emptyset$\;
 enqueue $Q$ with $v$\;
    \While{$Q$ \text{not empty}}{
        $u \leftarrow \text{ dequeue } Q$\;
        \For{$w \in \mathcal{N}(u)$}{
            \If{$w \not\in D$}{
                $p \leftarrow$ real number chosen u.a.r from $(0, 1)$\;
                \If{$p < \min \lbrace 1, \tfrac{\alpha}{|\mathcal{N}(u)|}\rbrace$}{
                $D \leftarrow D \cup \lbrace w \rbrace$\;
                enqueue $Q$ with $w$\;
                \lIf{$(u, w) \in E_{1}$}{$i \leftarrow 1$}
                \lIf{$(u, w) \in E_{2}$}{$j \leftarrow 1$}
                $V_{s} \leftarrow V_{s} \cup \lbrace (w,i,j) \rbrace$
                }
            }
        }
    }
\caption{Burn$(G = (V, E_{1}, E_{2}), v)$ \label{alg:alg2}}
\end{algorithm}

\begin{table}[b]
    \centering
	\caption{Same as \tref{tab:tab1}, but for the Twitter network extrapolated to $z = 1000$ using forest-fire densification. The layer density skewness $\delta_{z} = 29.6$, compared to $30.2$ in the original. Overlap is $\gamma = 0.45$, as before. \label{tab:tab2}}
		\begin{tabular}{c c c c c c c}
		\hline\hline
	    & $E$ & $E_{1}$ & $E_{2}$ & $E_{10}$ & $E_{01}$ & $E_{11}$\\
	  	\hline 
	  	$|\bullet|$ & $185,323,675$ & $6,139,522$ & $181,954,417$ & $3,369,258$ & $179,184,153$ & $2,770,264$\\
	    $z$ & $1000.27$ & $33.13$ & $982.1$ & $18.19$ & $967.1$ & $14.95$ \\
		\hline\hline
		\end{tabular}
\end{table}
We modify this algorithm in several ways to suit our framework; the authors of \cite{leskovec2007graph,kanade2016distance} consider single layer directed networks, whereas we require a process corresponding to undirected multiplexes. Further, \cite{leskovec2007graph} uses a geometric distribution to determine whether to burn through a particular edge. In contrast, we traverse edges adjacent to a node $v$ with probability $\tfrac{\alpha}{|\mathcal{N}(v)|}$, where $\mathcal{N}(v)$ is the set of neighbours of $v$, meaning $|\mathcal{N}(v)|$ provides the degree of $v$. Here, $\alpha$ is a parameter of the model determining the average number of edges to burn per node. For results in Fig. 3(d-f) of the main text, we set $\alpha = 1$, meaning that when a node $u$ is exploring the neighbours of a node $v$, it selects on average one with whom to connect. If $w \in \mathcal{N}(v)$, the edge type formed between $u$ and $w$ is determined by the type of edge $(v, w)$. In other words, if $u$ discovers $w$ via $v$, and $v$ is connected to $w$ via layer two and not layer one, then $u$ also connects to $w$ via layer two and not layer one. In the context of a growing Twitter multiplex, this corresponds to a form of cyclic closure~\cite{Rapoport1977Contribution,Kumpula2007Emergence}; if a user $u$ discovers a user $w$ via $v$, and $v$ has a follower / followee relationship with $w$ (not going so far as to form a mutual-mention bond), then $u$ is inclined to also form an inexpensive follower relationship with $w$. Although this argument may not apply in every instance, the assumption appears reasonable for a simple model of multiplex densification, and most importantly, preserves the balance of edge types in the multiplex (with density skewness maintained at $\delta_{z} \simeq 30.2$ even when the original network is extrapolated far beyond its original density, i.e., $z \gg 165.8$). 

Finally, to remain consistent with our approach to sparsification which preserves the network size $N$, we implement a variant of the forest-fire process that adds edges without adding nodes. In \cite{leskovec2007graph}, a new node randomly attaches to a node of the existing graph  (termed the ``ambassador'' node by Leskovec and coauthors), and then starts the forest-fire process at that node. In contrast, we begin each step by randomly selecting an existing node, and performing the forest-fire process from that node. As such, our implementation of the forest-fire process amounts to a randomised version of a graph traversal algorithm, such as a breadth first search (BFS). In fact, in the limiting case of $\alpha \rightarrow \infty$, the forest-fire process recovers the BFS. In this case, the node at which we start the forest-fire process eventually discovers every node in the graph. If this occurs at each step of our algorithm, eventually we will have a complete graph. If $\alpha = 0$, no new nodes are discovered and no edges added. As such, $\alpha$ parameterizes a Twitter user's tendency to recursively build its network. A consequence of this algorithm is that we're mostly adding short cycles to the network, with edges added after one hop forming 3-cycles, edges added after two hops forming 4-cycles, and so on\footnote{This is desirable since $(i)$ the formation of cycles is well motivated empirically \cite{Rapoport1977Contribution,Kumpula2007Emergence}, and $(ii)$, our goal in carrying out experiments on the Twitter dataset is to test our results beyond a configuration model setting, where networks are maximally random up to degree distribution, by construction. Clearly, one way this is achieved is through the addition of short cycles.}.
\begin{figure}
    \centering
    \includegraphics{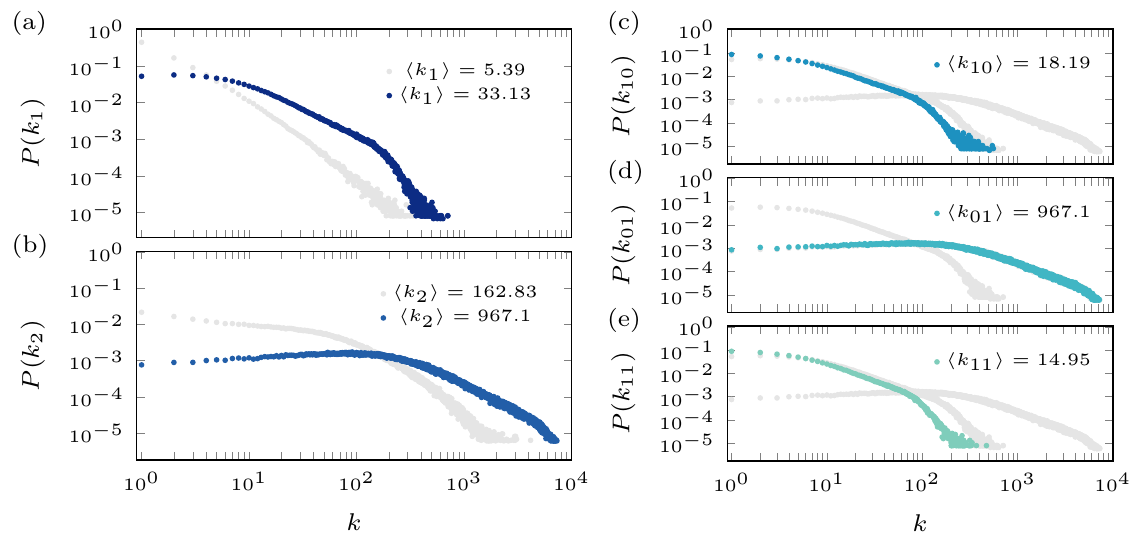}
    \caption{Degree distributions of the Twitter network extrapolated to $z = 1000$. Gray curves in (a) and (b) are the original distributions of $k_{1}$ and $k_{2}$ degrees [see \fref{fig:multiplex_figS6}(a) and (b)]. Gray curves in (c-e) are the blue curves in (a) and (b).}
    \label{fig:multiplex_figS10}
\end{figure}

\subsection{Emergence of unstable phase}

\begin{figure}
    \centering
    \hspace{-2mm}
    \vspace{-1mm}
    \includegraphics{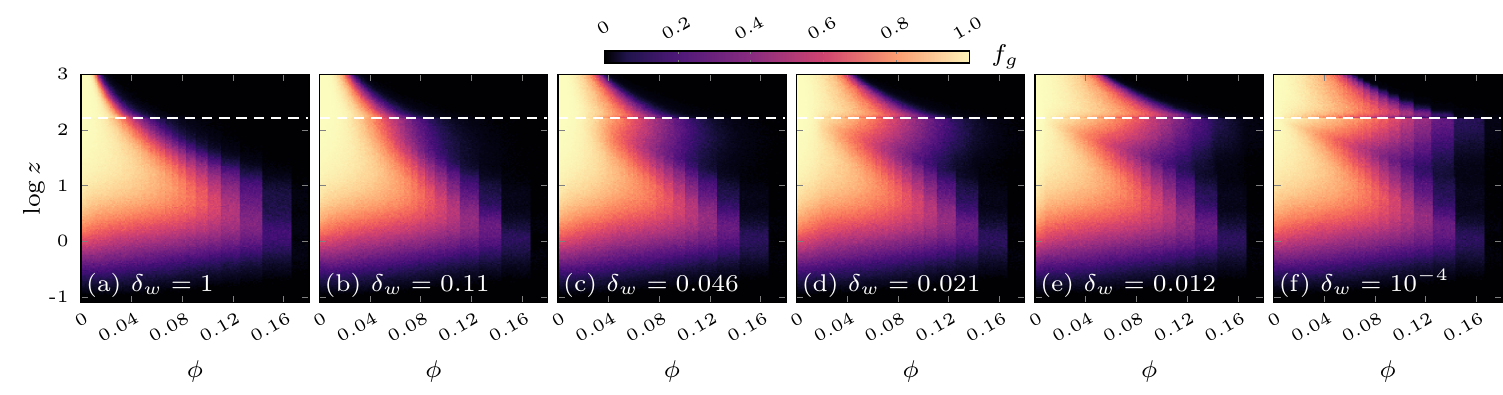}
    \caption{Emergence of a high-$z$ cascading phase due to increasing edge weight heterogeneity, or decreasing $\delta_{w}$, in an empirical Twitter network. Heat maps are the result of $10^{3}$ realisations of single node perturbation, and give the frequency $f_{g}$ of the emergence of global cascades. Here, this is an cascade whose steady state is of size $\rho_{g} = 5 \times 10^{-2}$. Alternatively, this is when a single node perturbation results in cascades of size $18500$ or higher. Horizontal dashed line is the original network density of $z = 165.8$.}
    \label{fig:multiplex_figS9}
\end{figure}

In this section, we demonstrate the emergence of a high-$z$ cascading phase in the empirical Twitter multiplex, using the weighted sum rule. In so doing, we summarise the arguments of previous sections in the language of real networks. Since this experiment involves varying the relative weights $w_{i}$ in each layer using $\delta_{w}$, which are assumed to be uniform here, the dynamics of the \textit{and} and \textit{or} threshold rule would be unchanged. In \fref{fig:multiplex_figS9}(a-f), we vary weight skewness $\delta_{w}$ from $1$, meaning weights are of equal strength in the mutual-mention and follower layers, to $10^{-4}$, such that the strength of interaction in the mutual-mention layer is $10^{4}$ times stronger than that in the follower layer. Weights are additive in overlapping links, which represent a substantial fraction $\gamma = 0.45$ of the mutual-mention layer. In each panel of \fref{fig:multiplex_figS9}, we explore a complete $(\phi, z)$ phase space, meaning we perform sparsification and densification to explore below and above the dashed horizontal line, respectively. This experiment is the empirical analogue of that presented in Fig. 2 of the main text, where we vary $\delta_{w}$ in a configuration model multiplex with log-normal degree distribution. The values of weight skewness $\delta_{w}$ in \fref{fig:multiplex_figS9}(a-f) correspond to weights $\textbf{w} = (1, 1)$, $(7, 0.8)$, $(13,0.6)$, $(19,0.4)$, $(25, 0.2)$ and $(30.7, 0.01)$, such that average weight across the entire multiplex gives $\langle w \rangle = 1$, for all plots.

When $\delta_{w} = 1$, and the mutual-mention network is of equal interaction strength to the follower network, a large overall $z$ ensures that global cascades are exponentially rare [\fref{fig:multiplex_figS9}(a)]. That is, if the threshold $\phi = 0.1$ for example, it suffices that the total average degree be $z > 30$ to ensure stability against global cascades. In this setting, no cascading phase is observed at high $z$, where the follower network overwhelms the influence of the mutual-mention network. In contrast, if weight heterogeneity is large, say $\delta_{w} = 0.012$ as in \fref{fig:multiplex_figS9}(e), then the multiplex has to reach a much higher average degree $z$ than in the previous case, before becoming stable to perturbations. This is due to the presence of a high-$z$ cascading phase. Now if $\phi = 0.1$, then the overall connectivity must surpass $z > 300$ before global cascades become exponentially rare. This can be understood by noting that the mutual-mention average degree lags behind that of the follower network. In the case of $\delta_{w} = 1$, by the time the mutual-mention network undergoes structural percolation, follower links are too abundant for influence from mutual-mention links to be perceptible. Taking the mutual-mention layer alone at $\phi = 0.1$ and $z = 5.4$, i.e., at its original value corresponding to the horizontal dashed line in (a), the network would be of ideal density to undergo global cascades. Since, however, the mutual-mention layer is coupled to the follower layer, and here they are of equal influence, $\delta_{w} = 1$, the nodes in the mutual-mention layer become highly stable. Thus, no cascades are observed at this point. In contrast, when $\delta_{w} = 0.012$, at that same point $(\phi = 0.1, z_{1} = 5.4)$ but in \fref{fig:multiplex_figS9}(e), mutual-mention links are of sufficient strength to overwhelm the influence from the follower network, despite being vastly outnumbered. Since the strong links are of ideal connectivity to trigger global cascades (percolating structurally and forming a giant component, but not too dense so as to stabilize adjacent nodes), global cascades emerge even from a single initial perturbation.

The above example suggests a straightforward manner to evaluate the susceptibility of an observed network to global cascades at an observed value of $z$. First, one attempts to determine whether the network is comprised of links of heterogeneous interaction strength. If so, and if this heterogeneity is substantial, even a high overall $z$ does not guarantee stability against external shocks. This is not the case when links are of homogeneous strength, where relatively low connectivity $z$ may be sufficient to suppress global cascades.

\bibliographystyle{unsrt}

\end{document}